\documentclass[conference]{IEEEtran}
\IEEEoverridecommandlockouts
\usepackage[bottom=0.75in,top=0.71in,left=0.7in,right=0.7in]{geometry}
\usepackage{amsmath}
\usepackage{amsfonts}
\usepackage{amssymb}
\usepackage{algpseudocode}
\usepackage{caption}
\usepackage{subfig}
\usepackage{mathtools}  
\usepackage{amsthm}
\usepackage{algorithm}
\usepackage{booktabs}
\usepackage{tabulary}
\usepackage{multirow}
\usepackage{footnote}
\usepackage[export]{adjustbox}
\usepackage{booktabs}
\usepackage{comment}
\newtheorem{remark}{Remark}
\usepackage{cases}
\usepackage{graphicx}
\usepackage{cite}
\usepackage{multicol}
\usepackage{xcolor}
\usepackage{graphicx}
\graphicspath{{./}{figures/}}
\newcommand\numeq[1]%
  {\stackrel{\scriptscriptstyle(\mkern-1.5mu#1\mkern-1.5mu)}{=}}
\usepackage{stfloats}
\usepackage{cuted}
\usepackage{lipsum}  

\usepackage{dsfont}
\usepackage{amsmath,amssymb}

\newtheorem{Prob}{\textbf{Problem}}

\newtheorem{theorem}{Theorem}

\usepackage{optidef}
\usepackage{amsmath}
\title{Hierarchical Agentic Mid-Band Agility for Reliable Low-Latency Space-Air-Ground Integrated V2X Networks}

 \author{ \text{Liu Cao},~\IEEEmembership{Member,~IEEE}, \text{Zisheng Gong},   \text{Lyutianyang Zhang},~\IEEEmembership{Member,~IEEE},\\ \text{Jinke Ren},~\IEEEmembership{Member,~IEEE}, \text{Jiguang He},~\IEEEmembership{Senior Member,~IEEE}, and \text{Shuguang Cui},~\IEEEmembership{Fellow,~IEEE}

 \thanks{Liu Cao and Zisheng Gong are with the Electrical and Electronic Engineering Program of both City University of Hong Kong (Dongguan), Dongguan 523808, China and City University of Hong Kong, Hong Kong (e-mail:\{liu.cao, zisheng.gong\}@cityu-dg.edu.cn).  

Lyutianyang Zhang is  with the School of Microelectronics and Communication Engineering, Chongqing
University, Chongqing 400044, China (email: zhanglyutianyang@cqu.edu.cn).
 
 Jinke Ren and Shuguang Cui are with the School of Science and Engineering, the Shenzhen Future Network of Intelligence Institute, and the Guangdong Provincial Key Laboratory of Future Networks of Intelligence, The Chinese University of Hong Kong (Shenzhen), Shenzhen, Guangdong
518172, China (email: \{jinkeren, shuguangcui\}@cuhk.edu.cn).  

Jiguang He is with the School of Computing and Information Technology, Great Bay University, Dongguan 523000, China (e-mail:
jiguang.he@gbu.edu.cn). (Corresponding author: Lyutianyang Zhang.)}

}

\begin{document}

\maketitle
\thispagestyle{empty}
\begin{abstract}
Upper-mid-band spectrum offers bandwidth opportunities for the integrated networks of ground vehicles, unmanned aerial vehicles (UAVs), and low Earth orbit (LEO) satellites to support 6G wide-area vehicle-to-everything (V2X) connectivity. However, V2X services' latency and reliability are sensitive to the mid-band selection due to frequency-dependent propagation and rain attenuation. This paper investigates environment-aware spectrum agility over candidate frequency range 3 (FR3) bands within 7.125-24.25~GHz, jointly with end-to-end (E2E) multi-path traffic splitting and transmit power control. Each data packet is partitioned into non-overlapping subflows over direct and UAV-assisted paths, with failed sky-link subflows selectively retransmitted subject to retransmission and freshness constraints. We formulate latency- and reliability-oriented problems, followed by a unified sequential optimization problem incorporating band-switching costs and average quality-of-service (QoS) violation budgets. To coordinate optimization decisions across heterogeneous timescales, we develop an agentic artificial intelligence (AI)-assisted hierarchical multi-agent reinforcement learning (AA-HMARL) framework. A high-level agentic supervisor integrates recurrent memory, analytical evaluation of candidate bands, and event-triggered replanning to coordinate slow environmental adaptation with fast channel and retransmission dynamics, while distributed ground vehicles and UAV agents optimize data splitting and power using constrained multi-agent proximal policy optimization (MAPPO). Simulation results demonstrate an adjustable reliability-latency trade-off. Particularly, AA-HMARL reduces the mean E2E completion latency and the mean number of retransmission rounds among successfully delivered updates by 17.8-25.5\% and 57.6-61.5\%, respectively, and reduces the mean age of information (AoI) by 12.7-14.6\% relative to fixed-band MAPPO across two traffic types.


\end{abstract}



\section{Introduction}
\label{sec:intro}


\IEEEPARstart{V}{ehicle}-to-everything (V2X) services increasingly require massive connectivity beyond the coverage of terrestrial infrastructure. Non-terrestrial networks (NTNs), including the satellite-access capabilities introduced in 3GPP Releases~17~\cite{3gpp_rel17_ntn} and~18~\cite{3gpp_rel18_ntn}, provide a complementary route toward wider-area connectivity.  Among different NTN platforms, low Earth orbit (LEO) satellites offer significantly reduced propagation delay compared with high-Earth orbit satellite systems and can provide wide-area connectivity beyond the coverage of terrestrial base stations\cite{wang2025environment}. Meanwhile, unmanned aerial vehicles (UAVs) provide flexible low-altitude relaying with favorable line-of-sight (LoS) propagation opportunities. These complementary advantages make the integrated vehicle-UAV-LEO network a promising architecture for extending reliable 6G V2X connectivity, especially when direct terrestrial infrastructure is unavailable, overloaded, or geographically constrained.

The emergence of mid-band spectrum further creates new opportunities for such integrated networks. The mid-band frequency range, spanning approximately $7.125$ to $24.25$ GHz and commonly referred to as frequency range 3 (FR3), provides a favorable balance between the wider bandwidth available at higher frequencies and the more desirable propagation characteristics of lower-frequency systems~\cite{11297179,rappaport2025spectrum,cui2025fr3}. However, their bandwidths and frequency-dependent propagation losses are different, and atmospheric and rain conditions can change their relative attractiveness. A wider-band candidate may therefore improve the transmission rate without providing the most favorable delivery reliability. This motivates \emph{FR3 spectrum agility}, where the operating frequency is adaptively selected according to the environment and service requirements. Such an adaptive mechanism is particularly relevant to vehicle-to-satellite (V2S) and UAV-to-satellite (U2S) links, whose long propagation distances and exposure to atmospheric conditions make their reliability and latency heavily dependent on the selected band.


At the same time, relying on a single V2S link makes the end-to-end (E2E) communication performance vulnerable to satellite visibility, channel degradation, and temporary sky-link interruptions. A natural alternative is to exploit the heterogeneity of the vehicle-UAV-LEO architecture by transmitting different parts of a packet through multiple parallel E2E paths, including direct V2S paths and UAV-assisted V2U-U2S relay paths\cite{10438999}. Unlike packet duplication, such multi-path traffic splitting assigns non-overlapping subflows to different paths, thereby allowing the traffic load to be jointly adapted to heterogeneous path rates and reliability conditions. However, a data packet can be reconstructed only after all active subflows have been successfully delivered, so retransmitting failed subflows may increase both E2E completion latency and information staleness. Age of information (AoI) has been increasingly used to characterize the timeliness of status updates in both vehicular and satellite networks. For instance, recent work has analyzed the AoI variation of LEO satellite-terrestrial uplinks under time-varying satellite channels, while AoI-aware resource allocation has been investigated in vehicular networks \cite{10879563}. Multi-path and retransmission-aware AoI designs have also been studied for LEO satellite routing, where constrained retransmission and traffic segmentation can improve information freshness under link disruptions \cite{11180797}. Nevertheless, these studies mainly focus on information-freshness-oriented routing or resource allocation. They do not jointly consider subflow-level selective retransmission, path-level traffic splitting, transmit-power control, and environment-aware FR3 band adaptation in integrated vehicle-UAV-LEO V2X networks\cite{11169451}. In the considered system, the retransmission process must be jointly constrained by both a maximum retransmission number and an AoI deadline, such that an excessively stale update is terminated rather than continuously retransmitted.

Resource allocation for UAV-assisted and integrated space-terrestrial networks has also attracted substantial attention\cite{mahboob2024ai_ntn,10795214,10622306}. Mean-field multi-agent reinforcement learning (MARL) has been employed for joint resource allocation in UAV-assisted V2X networks \cite{10689365}, while learning-based resource allocation has been investigated for integrated space-terrestrial intelligent transportation systems involving UAVs and LEO satellites \cite{10977731}. For integrated LEO and multi-UAV networks, federated multi-agent deep reinforcement learning has been further applied to jointly optimize UAV trajectories and communication resource allocation \cite{11124240,gong2026starmind}. For integrated satellite-terrestrial uplinks, joint two-tier association and resource management have been investigated to reduce transmission time \cite{10645749}. Channel prediction and MARL have also been combined for fair resource allocation and latency reduction in NTN uplinks \cite{10990315}. Moreover, cooperative transmission in space-air-ground integrated networks (SAGINs) has been addressed through multi-agent coordination \cite{10937131}. These studies demonstrate the potential of MARL for managing heterogeneous space-air-ground resources. However, their primary objectives are typically task offloading, coverage, throughput, energy efficiency, trajectory control, or computing-resource allocation, rather than the joint reliability-latency optimization considered in this paper. More importantly, conventional reinforcement-learning-based resource allocation generally operates at a single decision timescale, whereas the considered system inherently contains heterogeneous timescales: FR3 band selection should follow relatively slow environmental changes, while traffic splitting and transmit-power control must react to fast-timescale channel variations.

Recently, agentic AI has emerged as a promising technology for autonomous resource management in SAGINs. In contrast to conventional learning agents that directly map observations to actions, agentic AI frameworks incorporate additional capabilities such as environment perception, memory, reasoning, planning, tool invocation, and closed-loop adaptation. For example, an agentic AI framework based on semantic perception, intent-driven orchestration, and adaptive reinforcement learning has recently been proposed for SAGIN resource management \cite{11550346}. Such a framework is well suited to the considered problem because the slow-timescale FR3 spectrum decision requires not only instantaneous network observations but also historical environment-performance relations, model-based candidate-band evaluation, and event-triggered replanning. In contrast, the high-dimensional continuous control of path-level traffic splitting and transmit power is more naturally handled by MARL. This observation motivates a hierarchical design that combines agentic high-level spectrum planning with fast-timescale distributed resource allocation.


Motivated by these gaps, this paper addresses a coupled reliability-latency challenge: upper-mid-band FR3 band selection changes transmission rates and sky-link failure risks under frequency-dependent propagation and rain conditions, while any unfinished active subflow prevents complete update reconstruction. Consequently, FR3 spectrum adaptation must account for non-overlapping
traffic splitting, transmit power control, and freshness-limited
retransmissions together. We introduce a \emph{high-level agentic band supervisor} that integrates environmental perception, memory, model-tool evaluation, and event-triggered replanning to coordinate slow environmental adaptation with fast channel and
retransmission dynamics. We also introduce \emph{low-level distributed multi-agent proximal policy optimization (MAPPO) agents} to further control splitting and power per fast block, with failed sky-link subflows selectively retransmitted under retransmission and freshness limits. The main contributions are summarized as follows:

\begin{itemize}
    \item We develop an E2E multi-path-based integrated vehicle-UAV-LEO transmission architecture with environment-aware FR3 spectrum agility. This architecture links transmission latency and delivery reliability to the decisions made across transmission hops and parallel subflows.


    \item We formulate two complementary optimization problems corresponding to the latency and reliability, respectively, by jointly optimizing path-level traffic splitting, transmit-power allocation, and FR3 band selection. We further establish a unified normalized reliability-latency objective with explicit FR3 band-switching cost, thereby enabling controllable operation between latency-oriented and reliability-oriented regimes.

    \item We propose an agentic AI-assisted hierarchical MARL (AA-HMARL) framework tailored to the heterogeneous timescales of the considered problem. An agentic supervisor performs FR3 band selection at the slow timescale, while vehicle and UAV agents jointly optimize traffic splitting and transmit power at the fast timescale. Feasible traffic and power actions are enforced through constraint-aware action parameterization, while the reliability and latency requirements are incorporated through adaptive primal-dual learning.
\end{itemize}

The remainder of this paper presents the architecture in Sec.~\ref{secII}, the transmission and reliability models in Sec.~\ref{secIII}, and the formulated optimization problems in Sec.~\ref{secIV}. Section~\ref{secV} develops AA-HMARL, Sec.~\ref{secVI} reports the simulations, and Sec.~\ref{secVII} concludes the paper.

\section{System Architecture}
\label{secII}

\begin{figure*}[t]
    \centering
\includegraphics[width=.85\textwidth]{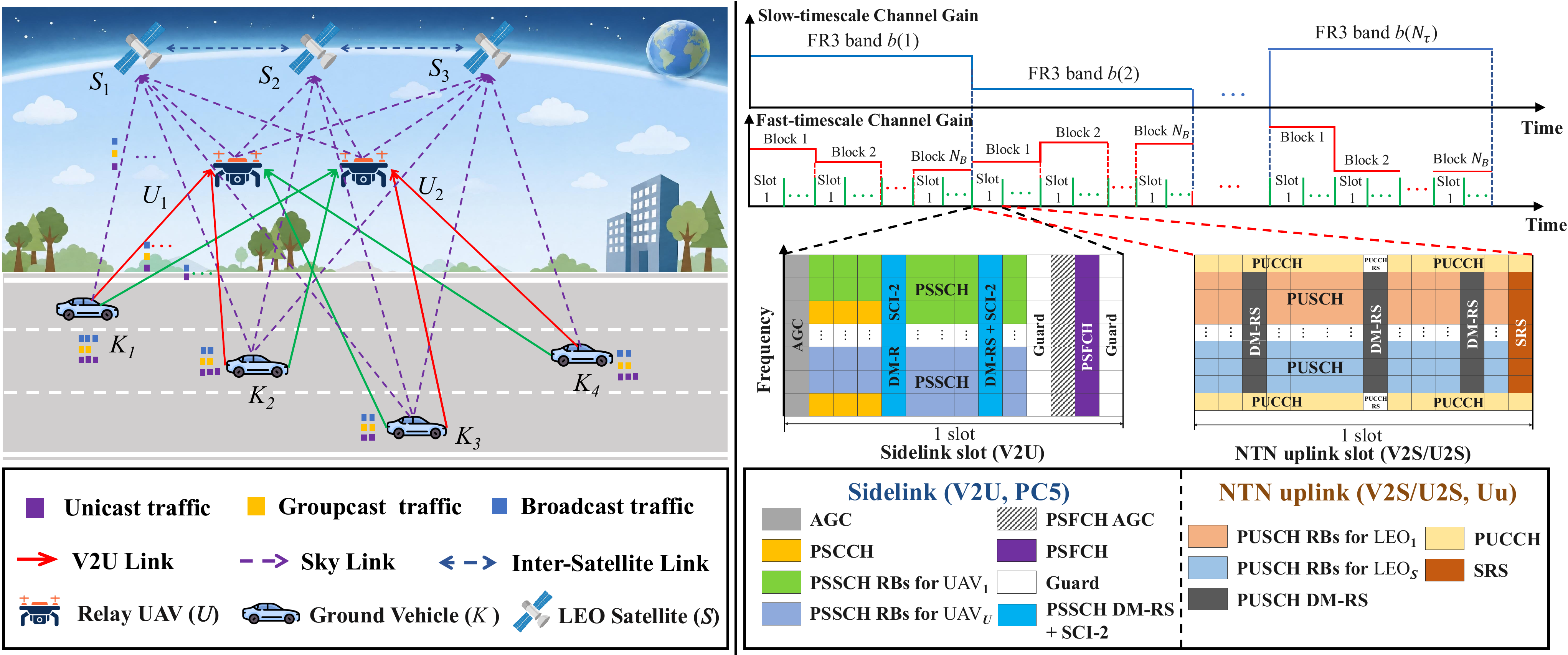}
    \vspace{-0.2cm}
    \caption{Heterogeneous E2E multi-path architecture with FR3 band agility for 6G vehicle-UAV-LEO networks.}
    \label{fig:E2Esysarch}
    \vspace{-0.6cm}
\end{figure*}

We consider a three-layer integrated vehicle-UAV-LEO satellite network, as illustrated in Fig.~1. The ground layer consists of a set of ground vehicles $\mathcal{K}\triangleq\{1,\ldots,K\}$ as source transmitters, the intermediate layer contains a set of relay UAVs $\mathcal{U}\triangleq\{1,\ldots,U\}$, and the space layer is composed of a set of LEO satellites $\mathcal{S}\triangleq\{1,\ldots,S\}$ as destination receivers.  Each ground vehicle can deliver $\mathcal{Q}\triangleq\{1,\ldots,Q\}$ heterogeneous V2X traffic types to the satellites\footnote{V2X traffic types can be categorized as unicast, groupcast, and broadcast per 3GPP standard \cite{3gpp.38.885}.}, either through a direct V2S link or indirectly through a two-hop V2U-U2S relay path\footnote{Subflows received by different LEO satellites are assumed to be forwarded via inter-satellite links to a common satellite for packet reassembly, while the associated inter-satellite forwarding and reassembly processes are beyond the scope of this work.}.  As a result, there are three types of wireless links that support data transmission from ground vehicles to satellites: V2U using the sidelink via the PC5 interface, and both V2S and U2S using the NTN uplink via the Uu interface. For vehicle $k \in  \mathcal{K}$, let $\mathcal{P}_{k}^{\rm D}$ denote the set of direct V2S paths and $\mathcal{P}_{k}^{\rm R}$ denote the set of UAV-assisted relay paths. A direct E2E path is represented by $p=(k,s) \in \mathcal{P}_{k}^{\rm D}$, which is a single V2S sky hop, and a relay path is represented by $p=(k,u,s) \in \mathcal{P}_{k}^{\rm R}$, consisting of a V2U access hop $(k,u)$ followed by a U2S sky hop $(u,s)$\footnote{The path index $p$ is also used to identify the corresponding V2S/U2S sky-link transmission, while the superscripts $\mathrm{V2U}$, $\mathrm{V2S}$, and $\mathrm{U2S}$ distinguish the constituent hop types when needed.}. The complete candidate E2E path set is therefore given by $\mathcal{P}_{k}
    =
    \mathcal{P}_{k}^{\rm D}
    \cup
    \mathcal{P}_{k}^{\rm R}$. 
    
    The LEO satellites are assumed to support a finite set of candidate upper-mid-band frequencies within the FR3 spectrum. Let $\mathcal{B}\triangleq\{1,\ldots,B\}$ denote the candidate FR3 band set, and $\mathcal{S}_b\subseteq\mathcal{S}$ denote the set of LEO satellites available over band $b \in \mathcal{B}$.  Each ground vehicle is allowed to simultaneously communicate with multiple candidate UAVs and LEO satellites to exploit the heterogeneous transmission opportunities provided by the three-layer architecture. Thus, the heterogeneous V2X traffic generated by each ground vehicle is split into \emph{different} subflows and concurrently transmitted over multiple E2E paths. Meanwhile, transmissions from the same transmitter to different receivers occupy orthogonal time-frequency resource blocks (RBs), whereas different transmitters serving the same receiver may reuse RBs and cause co-channel interference. Thus, the outgoing subflows from the same ground vehicle or UAV cause no intra-transmitter cross-subflow interference, while different ground vehicles/UAVs serving the same UAV/LEO satellite may interfere with each other. As the V2U access and V2S/U2S sky-links use different bands, cross-tier interference is neglected.

We consider a block-based transmission structure under a block-fading model, where the instantaneous channel gain remains approximately constant within each fast block and varies across blocks due to mobility and time-varying propagation conditions. At the beginning of each fast block, each ground vehicle determines E2E traffic-splitting ratios for different traffic and V2U/V2S transmit powers for different subflows, whereas each UAV allocates U2S transmit power among the buffered subflows. During selective sky-link retransmissions, the original path assignments and splitting ratios remain unchanged, while the corresponding transmit powers can be updated. As illustrated in Fig.~\ref{fig:E2Esysarch}, the representative sidelink and NTN uplink frame structures \cite{3gpp.38.885,3gpp38300} are organized within a common fast-block scheduling timeline. In the V2U sidelink slot, different physical sidelink shared channel (PSSCH) resource blocks are assigned to subflows for V2U, while the physical sidelink control channel (PSCCH) carries first-stage sidelink control information (SCI) for V2U. The PSSCH carries the corresponding V2U subflow data and second-stage SCI. Similarly, in the NTN uplink slot, distinct PUSCH resource blocks accommodate V2S/U2S subflow transmissions toward different LEO satellites. The physical uplink shared channel (PUSCH) carries the subflow data, whereas the physical uplink control channel (PUCCH) supports uplink control information (UCI). 

In this paper, the V2U access band remains fixed, reflecting relatively stable terrestrial-to-low-altitude propagation, whereas V2S/U2S sky links exploit environment-aware FR3 spectrum agility. The selected FR3 sky-link band $b(\tau)\in\mathcal{B}$ is normally retained for $N_{\rm B}$ fast blocks, where $\tau$ is the high-level decision index. Let $t_\tau$ denote the starting fast-block index of the $\tau$-th high-level decision, and $d_\tau$ denote its actual duration in fast blocks. The corresponding block set is defined as $\mathcal{N}_\tau \triangleq \{t_\tau,\ldots,t_\tau+d_\tau-1\}$, where $d_\tau=|\mathcal{N}_\tau|$. However, this decision may terminate earlier and a new decision is triggered if a significant environmental or quality-of-service (QoS) variation is detected. Due to the long propagation distance and environment-dependent attenuation of the V2S and U2S links, a sky-link subflow may fail during transmission. We therefore employ a selective retransmission mechanism at the subflow level: If one subflow fails while other concurrently transmitted subflows are successfully delivered, only the failed subflow is retransmitted, whereas all successfully received subflows are retained and do not participate in subsequent retransmissions. The path assignment of each subflow remains unchanged throughout the retransmission.

\section{System Model}
\label{secIII}



\subsection{V2U Transmission Model} 



Suppose each ground vehicle is equipped with an $N_{\rm V}$-element uniform linear array (ULA), whereas each UAV employs an $N_x\times N_y$ uniform planar array (UPA), with $N_{\rm U}=N_xN_y$. Let $\mathbf{p}_k(n)\in\mathbb{R}^{3}$ and $\mathbf{q}_u(n)\in\mathbb{R}^{3}$ denote the positions of vehicle $k$ and UAV $u$, respectively, in block $n$. Their three-dimensional distance is $d_{k,u}(n) = \left\| \mathbf{p}_k(n)-\mathbf{q}_u(n) \right\|_2$. 
Since the V2U links operate over a fixed access band with carrier frequency $f_{\rm A}$, their large-scale path loss is modeled as $L_{k,u}^{\rm V2U}(n) =
20\log_{10} \left( \frac{4\pi f_{\rm A}d_0}{c} \right) + 10\eta_{\rm V2U} \log_{10} \left( \frac{d_{k,u}(n)}{d_0} \right) 
+ X_{k,u}^{\sigma}(n)$, 
where $d_0$ is a reference distance, $\eta_{\rm V2U}$ is the path-loss exponent, $X_{k,u}^{\sigma}(n)$ represents the shadow fading, and $c$ is the speed of light. Accordingly, the large-scale channel gain is $
\beta_{k,u}^{\rm V2U}(n) = 10^{-L_{k,u}^{\rm V2U}(n)/10}$. 
Under the block fading assumption, the V2U multiple-input multiple output (MIMO) channel from vehicle $k$ to UAV $u$ is modeled as \begin{equation} 
\small
\begin{split} 
\mathbf{H}_{k,u}^{\rm V2U}(n) =&\, \sqrt{\beta_{k,u}^{\rm V2U}(n)} \rho_{k,u}(n) \mathbf{a}_{\rm U} \left( \theta_{k,u}^{\rm AoA}(n), \phi_{k,u}^{\rm AoA}(n) \right) \\ &\times \mathbf{a}_{\rm V}^{H} \left( \phi_{k,u}^{\rm AoD}(n) \right), 
\end{split} 
\end{equation} 
where $\rho_{k,u}(n)$ denotes the small-scale fading coefficient, while $\mathbf{a}_{\rm U}(\cdot)$ and $\mathbf{a}_{\rm V}(\cdot)$ denote the UAV-side UPA and vehicle-side ULA steering vectors, respectively. Let $\mathbf{v}_{k,u}(n)$ and $\mathbf{w}_{k,u}(n)$ denote the normalized transmit beamformer and receive combiner, respectively. If different vehicles reuse the same V2U resource toward UAV $u$, the post-combining signal to interference plus noise ratio (SINR) is \begin{equation} 
\small
\begin{split} 
\gamma_{k,u}^{\rm V2U}(n) = \frac{ P_{k,u}^{\rm V2U}(n) \left| \mathbf{w}_{k,u}^{H}(n) \mathbf{H}_{k,u}^{\rm V2U}(n) \mathbf{v}_{k,u}(n) \right|^2 }{ \displaystyle \sum_{\substack{k'\in\mathcal{K}\\k'\neq k}} P_{k',u}^{\rm V2U}(n) \left| \mathbf{w}_{k,u}^{H}(n) \mathbf{H}_{k',u}^{\rm V2U}(n) \mathbf{v}_{k',u}(n) \right|^2 + N_0W_{\rm A} }, 
\end{split} 
\end{equation} 
where $N_0$ denotes the noise power spectral density and $W_{\rm A}$ is the V2U access bandwidth. Hence, the corresponding achievable data rate is given by
\begin{equation} R_{k,u}^{\rm V2U}(n) = W_{\rm A} \log_2 \left( 1+\gamma_{k,u}^{\rm V2U}(n) \right). \end{equation} 


\subsection{Environment-Aware FR3 Mid-Band Sky-Link Model} 
We then characterize the corresponding sky-link under the candidate FR3 band. We denote the slow-varying environment state in the band-selection interval $\tau$ as $\mathbf e(\tau)=[R(\tau),I_{\rm eff}(\tau)]$, where $R(\tau)$ is the rain intensity, and $I_{\rm eff}(\tau)$ is the effective ionospheric stress. In particular, vertical total electron content (VTEC) represents the integrated electron density along the vertical ionospheric column. Let $\Delta\mathrm{VTEC}$ and $\dot{\mathrm{VTEC}}$ denote its deviation from a background reference and its temporal variation rate, respectively \cite{pi1997monitoring,juan2018aatr}. An ionospheric estimator combines these quantities into $I_{\mathrm{eff}}= \omega_1[\Delta\mathrm{VTEC}]^{+} +\omega_2|\dot{\mathrm{VTEC}}|$, where $\omega_1$ and $\omega_2$ are calibrated weights.
Let $A_{p,b}^{\rm env}(n)$ denote the environment-induced attenuation experienced by the sky hop of path $p$ operating over band $b$, which can be modeled as
\begin{equation} 
\small
A_{p,b}^{\rm env}(n) = k_bR^{\eta_b}(\tau) \rho_{p}^{\rm rain}(n) + \gamma_b^{\rm gas} \rho_{p}^{\rm gas}(n) + \chi_{p}c_bI_{\rm eff}(\tau) + \Delta m_{p,b}, 
\label{eq:total}
\end{equation} 
where $k_b$ and $\eta_b$ characterize the frequency-dependent rain attenuation, $\rho_{p}^{\rm rain}(n)$ is the effective rain-path length, $\gamma_b^{\rm gas}$ and $\rho_{p}^{\rm gas}(n)$ characterize gaseous absorption, and $\Delta m_{p,b}$ is an uncertainty margin. The ionospheric coefficient satisfies $c_b=c_0(f_0/f_b)^{\nu}, \nu>0$,
with $\chi_p=1$ for sky links. These coefficients capture the frequency-dependent environmental losses across FR3. Let $g_{p,b}^{0}(n)$ denote the clear-weather-equivalent power gain of sky link $p$ over FR3 band $b$, including the antenna gains, free-space propagation loss, and implementation losses.  After accounting for the environment-induced attenuation in Eq. \eqref{eq:total}, the effective power gain is
\begin{equation}
g_{p,b}(n)
=
g_{p,b}^{0}(n)
10^{-A_{p,b}^{\rm env}(n)/10}.
\end{equation}
For a feasible E2E path $p\in\mathcal{P}_k(n)$, we define the transmit power of its associated sky-hop transmission as
\begin{equation}
 P_p^{\rm sky}(n)=
 \begin{cases}
 P_{k,s}^{\rm V2S}(n),&p=(k,s)\in\mathcal{P}_{k}^{\rm D},\\
 P_{k,u,s}^{\rm U2S}(n),&p=(k,u,s)\in\mathcal{P}_{k}^{\rm R}.
 \end{cases}
 \label{eq:path-sky-power}
\end{equation}

Let $I_{p,b}^{\rm sky}(n;\mathbf{P})$ denote the aggregated co-channel interference generated by other concurrent V2S/U2S transmissions that reuse the same time-frequency resource and are directed toward the same LEO satellite as the sky-hop transmission associated with path $p$. The resulting sky-link SINR is
\begin{equation}
\gamma_{p,b}^{\rm sky}(n;\mathbf{P})
=
\frac{
P_{p}^{\rm sky}(n)g_{p,b}(n)
}{
I_{p,b}^{\rm sky}(n;\mathbf{P})
+
N_0W_b
},
\label{eq:path-sky-sinr}
\end{equation}
where $I_{p,b}^{\rm sky}(n;\mathbf{P})$ is determined by the allocated transmit power and effective channel gains of the concurrent co-channel transmissions. Hence, the sky-link SINR is coupled with the network-wide transmit-power set $\mathbf{P}(n)
=
\left[\{P_{k,u}^{\rm V2U}(n)\},
\{P_{p}^{\rm sky}(n)\}\right]$. Accordingly, the achievable rate of the sky-hop transmission associated with path $p$ is given by
\begin{equation}
R_{p,b}^{\rm sky}(n)
=
W_b
\log_2
\left(
1+\gamma_{p,b}^{\rm sky}(n)
\right),
\label{eq:path-sky-rate}
\end{equation}
where $W_b$ is the effective bandwidth assigned to a scheduled sky-link transmission under band $b$.





\subsection{E2E Initial Transmission Latency Model} 

Let $\Xi_{k,q}(n)$ be the payload bytes belonging to the status update of traffic type $q$ for vehicle $k$ scheduled in block $n$. Then the subflow size of traffic type $q$ assigned to path $p$ is given by $\Xi_{p,q}(n)
=
\alpha_{p,q}(n)\Xi_{k,q}(n)$,
where $0\leq\alpha_{p,q}(n)\leq1, $ and $ \sum_{p\in\mathcal{P}_{k}(n)} \alpha_{p,q}(n)=1
$. In particular, the active path set is defined as $\mathcal{A}_{k,q}(n) = \left\{ p\in\mathcal{P}_{k}(n): \alpha_{p,q}(n)>0 \right\}$. To distinguish the initial transmission from subsequent retransmissions, let $a\in\mathbb{Z}_{+}$ denote the sky-link transmission attempt index, where $a=1$ corresponds to the initial transmission and $a\geq2$ corresponds to the $(a-1)$-th retransmission. We first characterize the latency of the initial transmission attempt $a=1$. The E2E latency of a direct V2S path $p \in \mathcal{P}_{k}^{\rm D}(n)$ under the initial sky-link transmission attempt is given by
\begin{equation}
L_{p,q}^{\rm D,(1)}(n)
=
\frac{\Xi_{p,q}(n)}
{R_{p,b}^{\rm sky}(n)}
+
d_{k,s}^{\rm prop}(n)
+
d_s^{\rm proc},
\label{eq:direct-initial-latency}
\end{equation}
where $d^{\rm prop}$ and $d^{\rm proc}$ are the propagation and processing delays. For a relay path $p \in \mathcal{P}_{k}^{\rm R}(n)$,  the total subflow size of traffic type $q$ transmitted from ground vehicle $k$ to UAV $u$ is $\Xi_{k,u,q}^{\rm V2U}(n)
=
\sum_{s: p\in \mathcal{A}_{k,q}(n)}
\Xi_{p,q}(n)$.
Under the store-and-forward relay assumption, the V2U access latency is given by
\begin{equation}
L_{k,u,q}^{\rm V2U}(n)
=
\frac{
\Xi_{k,u,q}^{\rm V2U}(n)
}{
R_{k,u}^{\rm V2U}(n)
}
+
d_{k,u}^{\rm prop}(n)
+
d_u^{\rm proc}.
\label{eq:v2u-access-latency}
\end{equation}

Let $B_{p,q}(n)$ denote the amount of unresolved type $q$ traffic buffered at UAV $u$ for subsequent transmission toward satellite $s$ at the beginning of block $n$. Let $A_{p,q}(n)$ denote the newly admitted traffic that has completed its V2U delivery and enters this buffer during block $n$, which is determined by the traffic-splitting decision $\alpha_{p,q}(n)$ through $\Xi_{p,q}(n)$. Let $S_{p,q}(n)$ denote the amount of buffered traffic successfully delivered over the U2S sky hop associated with path $p$ during block $n$, which satisfies
$0
\leq
S_{p,q}(n)
\leq
R_{p,b}^{\rm sky}(n)T_{\rm blk}$,
where $T_{\rm blk}$ is the block duration. In particular, $S_{p,q}(n)=0$ when the scheduled U2S transmission attempt fails. The UAV-side buffer is given by 
\begin{equation}
B_{p,q}(n+1)
=
\left[
B_{p,q}(n)
-
S_{p,q}(n)
\right]^+
+
A_{p,q}(n),
\label{eq:uav-buffer-evolution}
\end{equation}
where $[x]^+\triangleq\max\{x,0\}$. Using a fluid-service approximation\cite{914515}, the corresponding UAV-side queueing delay is modeled as
$d_{p,q}^{\rm que}(n)
=
\frac{
B_{p,q}(n)
}{
R_{p,b}^{\rm sky}(n)
}$. After decoding and buffering at UAV $u$, the subflow
associated with relay path $p$ experiences the
initial U2S latency
\begin{equation}
L_{p,q}^{\rm U2S,(1)}(n)
=
d_{p,q}^{\rm que}(n)
+
\frac{
\Xi_{p,q}(n)
}{
R_{p,b}^{\rm sky}(n)
}
+
d_{u,s}^{\rm prop}(n)
+
d_s^{\rm proc}.
\label{eq:u2s-initial-latency}
\end{equation}
As a result, the E2E latency of relay path $p \in \mathcal{P}_{k}^{\rm R}$ under its initial sky-link transmission attempt is given by
\begin{equation}
L_{p,q}^{\rm R,(1)}(n)
=
L_{k,u,q}^{\rm V2U}(n)
+
L_{p,q}^{\rm U2S,(1)}(n).
\label{eq:relay-initial-latency}
\end{equation}

\begin{theorem}
\label{thm:latency-splitting}
Assume a nonempty feasible path set $\mathcal P_k$, fixed positive link rates,
fixed path-latency offsets $d_p$, and $\Xi_{k,q}>0$. Define
\begin{equation}
\small
\begin{aligned}
C_k\triangleq\sum_{p\in\mathcal P_k^{\rm D}}R_{p,b}^{\rm sky}+\sum_{u: p\in \mathcal{P}_{k}^{\rm R}}
\frac{R_{k,u}^{\rm V2U}\displaystyle\sum_{s:p\in\mathcal P_{k}^{\rm R}}R_{p,b}^{\rm sky}}
{R_{k,u}^{\rm V2U}+\displaystyle\sum_{s:p\in\mathcal P_{k}^{\rm R}}R_{p,b}^{\rm sky}}.
\end{aligned}
\label{eq:effective-aggregate-rate}
\end{equation}
Let
$\ell_{p,q}^{(1)}$ denote the initial latency of path $p \in \mathcal{P}_{k}$. The initial E2E completion latency
$D_{k,q}^{(1)}=\max_{p\in\mathcal A_{k,q}}\ell_{p,q}^{(1)}$
satisfies
\begin{equation}
D_{k,q}^{(1)}\geq \min_{p\in\mathcal P_k}d_p+\frac{\Xi_{k,q}}{C_k}.
\label{eq:initial-latency-bound}
\end{equation}
\end{theorem}
\begin{proof}
Let $d_{\min}=\min_{p\in\mathcal P_k}d_p$ and
$T=(D_{k,q}^{(1)}-d_{\min})/\Xi_{k,q}\geq0$.
For the direct path $p\in\mathcal P_k^{\rm D}$, it satisfies
$\alpha_{p,q}\leq T R_{p,b}^{\rm sky}$. For the relay path $p\in\mathcal P_k^{\rm R}$, write
$a_u=\sum_{s: p\in\mathcal P_{k}^{\rm R}}\alpha_{p,q}$ and
$B_u=\sum_{s:p\in\mathcal P_{k}^{\rm R}}R_{p,b}^{\rm sky}$.
If $a_u>0$, the latency of each path in the relay path group implies $\frac{a_u}{R_{k,u}^{\rm V2U}}+
\frac{\alpha_{p,q}}{R_{p,b}^{\rm sky}}\leq T$.
 Multiplying by $R_{p,b}^{\rm sky}$,
summing over the relay path group gives
$a_u\leq B_u\left(T-\frac{a_u}{R_{k,u}^{\rm V2U}}\right)$, and rearranging further gives $
a_u\leq T\frac{R_{k,u}^{\rm V2U}B_u}{R_{k,u}^{\rm V2U}+B_u}$.
Summing the direct path and relay path group bounds yields $1=\sum_{p\in\mathcal P_k^{\rm D}}\alpha_{p,q}
+\sum_{u:p \in \mathcal P_{k}^{\rm R}}a_u
\leq T C_k$.
Substituting the definition of $T$ proves Eq. \eqref{eq:initial-latency-bound}.
\end{proof}
Theorem 1 shows that the initial E2E completion latency cannot be reduced
below the derived lower bound by any feasible traffic-splitting
strategy.

\subsection{E2E Sky-Link Reliability Model} 


The FR3 V2S and U2S sky-links are subject to environment-induced transmission interruption. 
Once a relay subflow has been successfully delivered from ground vehicle $k$ to UAV $u$, it remains buffered at the UAV until the corresponding U2S delivery is completed or terminated. Hence, a U2S failure does not trigger a repeated V2U transmission. Let $\delta_{p,q}^{(a)}\in\{0,1\}$ denote the success indicator of the $a$-th sky-link transmission attempt, where $\delta_{p,q}^{(a)}=1$ indicates successful delivery and $\delta_{p,q}^{(a)}=0$ otherwise. Denote $z_{p,q}^{(1)}=1$ as the unresolved subflow indicator, and 
$z_{p,q}^{(a+1)} = z_{p,q}^{(a)} \left( 1-\delta_{p,q}^{(a)} \right)$.
Therefore, once a subflow is successfully received, $z_{p,q}^{(a+1)} = 0$, which indicates the subflow consumes no further transmission resources.  Let $R_{p,q}^{\max}$ denote the maximum number of retransmissions allowed for path $p$ due to multiple retransmission failure. Then the corresponding subflow data will be dropped if $z_{p,q}^{(R_{p,q}^{\max}+1)}=1$.

In addition, the retransmission limit is also constrained by the freshness deadline
of the served packet. To characterize receiver-side
information freshness, let $g_{k,q}^{\mathrm{last}}(t)$ denote the generation time of the freshest completely delivered packet of traffic type $q$ from vehicle $k$ available at the satellite
layer at time $t$. The instantaneous AoI
is defined as
$\Delta_{k,q}(t)=t-g_{k,q}^{\mathrm{last}}(t)$.
As one packet is partitioned into different non-overlapping subflows, reception of only a subset of subflows does not reset the AoI. Let $Y_{k,q}(n)\in\{0,1\}$ denote whether a complete packet fresher than the currently available one is successfully delivered during block $n$, which is incorporated at the end of the block for the AoI evolution. Let $D_{k,q}^{\rm sys}(n)$ denote the age of the freshest newly delivered packet, measured from its generation time to the end of block $n$. Accordingly, the AoI at the end of block evolves as
\begin{equation} 
\small
\Delta_{k,q}(n+1) = Y_{k,q}(n) D_{k,q}^{\rm sys}(n) + \left( 1-Y_{k,q}(n) \right) \left( \Delta_{k,q}(n)+T_{\rm blk} \right), 
\end{equation} 
where $T_{\rm blk}$ is the block duration. The long-term average AoI of traffic type $q$ for vehicle $k$ is 
\begin{equation} 
\overline{\Delta}_{k,q} = \limsup_{N\rightarrow\infty} \frac{1}{N} \sum_{n=1}^{N} \mathbb{E} \left[ \Delta_{k,q}(n) \right]. 
\end{equation} 

Let $a_{k,q}^{\rm pkt}(n)$ denote the age of the currently served packet at the beginning of block $n$, and $\Delta_{k,q}^{\max}$ denote its freshness deadline. The AoI-limited retransmission number is defined as
\begin{equation}
\small
\begin{split}
 R_{p,q}^{\rm AoI}(n)=\max\Big\{r\in\mathbb Z_{\geq0}:\;&a_{k,q}^{\rm pkt}(n)+ L_{p,q}^{\rm acc}(n)\\
 &+\sum_{a=1}^{r+1}L_{p,q}^{\rm sky,(a)}(n)
 \leq\Delta_{k,q}^{\max}\Big\},
\end{split}
\end{equation}
where $L_{p,q}^{\rm acc}(n)$ is the one-time V2U access latency and $ L_{p,q}^{\rm acc}(n) = 0$ for a direct V2S path.  $L_{p,q}^{\rm sky,(a)}(n)$ is the latency of the $a$-th sky-link transmission attempt. Then the effective retransmission limit is given by 
$R_{p,q}^{\rm eff}(n)=\min\{R_{p,q}^{\max},R_{p,q}^{\rm AoI}(n)\}$.  The realized accumulated transmission latency of subflow $p$ is expressed as
\begin{equation} 
\begin{split} 
D_{p,q}(n) = L_{p,q}^{\rm acc}(n) + \sum_{a=1}^{R_{p,q}^{\rm eff}(n)+1} z_{p,q}^{(a)} L_{p,q}^{\rm sky,(a)}(n). 
\end{split} 
\end{equation} 
As a packet is successfully delivered only when all active subflows are received, its E2E completion latency, conditioned on successful delivery, is determined by the last completed subflow, as
\begin{equation} 
\begin{split} 
D_{k,q}(n) = \max_{p\in\mathcal{A}_{k,q}(n)} D_{p,q}(n). 
\end{split} 
\end{equation} 

To characterize sky-link reliability, 
let $\gamma_{p,b,q}^{\rm req}$ denote the required SINR threshold for traffic type $q$. Then the corresponding SINR deficit is defined as $Z_{p,b,q}(n)
=
\gamma_{p,b,q}^{\rm req}
-
\gamma_{p,b}^{\rm sky}(n)$,
which gives a risk score 
$
s_{p,b,q}^{\rm risk}(n)
=
\Phi
\left(
\frac{
Z_{p,b,q}(n)
}{
\sigma_{p,b,q}^{Z}
}
\right)$,
where $\Phi(\cdot)$ denotes the standard Gaussian cumulative distribution function (CDF) and $\sigma_{p,b,q}^{Z}>0$ is a calibration parameter. As a result, the conservative upper envelope of the single-attempt sky-link outage probability is 
\begin{equation} 
\begin{split} 
p_{p,b,q}(n)
=
\left[
s_{p,b,q}^{\rm risk}(n)
+
\beta_{\rm r}
\sigma_{p,b,q}^{p}(n)
\right]_{0}^{1},
\end{split} 
\end{equation} 
where $[x]_{0}^{1}\triangleq\min\{1,\max\{0,x\}\}$, $\sigma_{p,b,q}^{p}$ represents the uncertainty radius, and $\beta_{\rm r}\geq0$ controls the conservativeness of the risk estimate. Suppose the $a$-th sky-link transmission attempt associated with path $p$ has the conservative outage envelope $p_{p,b_a,q}^{(a)}$. Each retransmission attempt is assumed to be independent across blocks. As $p_{p,b_a,q}^{(a)}$ upper-bounds the outage probability of the $a$-th retransmission attempt, a conservative lower bound on the probability that subflow $p$ is successfully delivered within its effective retransmission limit is expressed as
\begin{equation}
Q_{p,q}(n)
=
1-
\prod_{a=1}^{R_{p,q}^{\rm eff}(n)+1} p_{p,b_a,q}^{(a)}.
\label{eq:path-reliability}
\end{equation}
However, the sky links may remain statistically correlated across different paths because they experience a common large-scale rain and ionospheric state. Hence, we construct a unified conservative E2E reliability lower bound,  given by
\begin{equation} 
\small
\begin{split} 
\mathcal{R}_{k,q}(n) = \left[ 1- \sum_{p\in\mathcal{A}_{k,q}(n)} \left( 1-Q_{p,q}(n) \right) \right]^+.
\end{split} 
\end{equation}

\section{E2E Reliability and Latency Optimization}
\label{secIV}

In this section, we formulate two complementary optimization problems for the proposed vehicle-UAV-LEO transmission architecture, corresponding to the latency-oriented and reliability-oriented operating points, respectively. 


Since a status update may be dropped once it exceeds its effective retransmission limit, we define its conditional expected E2E completion latency as $\overline{D}_{k,q}(n)
    =
    \mathbb{E}
    \left[
    D_{k,q}(n)
    \mid
    Y_{k,q}(n)=1
    \right]$,
where $Y_{k,q}(n)=1$ indicates that all constituent subflows of the packet are successfully delivered and reassembled. For heterogeneous traffic types, let $\omega_q\geq0$ denote the service-priority weight, where $\sum_{q\in\mathcal{Q}}\omega_q=1$. The average E2E latency and reliability over interval $\tau$ are respectively defined as
\begin{equation}
\small
\begin{aligned}
    \overline{D}_{\rm E2E}(\tau)
    =
    \frac{1}{K|\mathcal{N}_{\tau}|}
    \sum_{n\in\mathcal{N}_{\tau}}
    \sum_{k\in\mathcal{K}}
    \sum_{q\in\mathcal{Q}}
    \omega_q
    \overline{D}_{k,q}(n),
    \end{aligned}
\end{equation}
\begin{equation}
\small
\begin{aligned}
    \overline{\mathcal{R}}_{\rm E2E}(\tau)
    =
    \frac{1}{K|\mathcal{N}_{\tau}|}
    \sum_{n\in\mathcal{N}_{\tau}}
    \sum_{k\in\mathcal{K}}
    \sum_{q\in\mathcal{Q}}
    \omega_q
    \mathcal{R}_{k,q}(n).
        \end{aligned}
\end{equation}

For notational convenience, the fast-timescale
traffic-splitting variables are defined as
$\boldsymbol{\alpha}(n)
=\left[\{\alpha_{p,q}(n)\}\right]$. 
The slow-timescale sky-band decision is denoted by $b(\tau)\in\mathcal{B}$. Our first objective is to minimize the E2E completion latency while guaranteeing a prescribed reliability level, which is formulated as follows.

\begin{Prob}[E2E Completion Latency Optimization]
\label{Minimum Latency}
\small
\begin{subequations}
\label{eq:latency-objective}
\begin{align}
&\min_{\substack{
\{\boldsymbol{\alpha}(n),\,\mathbf{P}(n)\}_{n\in\mathcal{N}_{\tau}},\,
b(\tau)}}
\quad \overline{D}_{\mathrm{E2E}}(\tau)
\tag{\theparentequation}
\\[2pt]
&\mathrm{s.t.}\quad \mathrm{C1:}~
0\leq \alpha_{p,q}(n)\leq 1,
\quad
\forall p\in\mathcal{P}_{k}(n;b(\tau)),\,k,q,n,
\label{eq:latency-c1}
\\
&\phantom{\mathrm{s.t.}\quad}\mathrm{C2:}~
\sum_{p\in\mathcal{P}_{k}(n;b(\tau))}
\alpha_{p,q}(n)=1,
\quad \forall k,q,n,
\label{eq:latency-c2}
\\
&\phantom{\mathrm{s.t.}\quad}\mathrm{C3:}~
\sum_{u\in\mathcal{U}}P_{k,u}^{\mathrm{V2U}}(n)
+
\sum_{p\in\mathcal{P}_{k}^{\rm D}(n;b(\tau))}
P_{p}^{\rm sky}(n)
\leq P_{k}^{\max},
\quad \forall k,n,
\label{eq:latency-c3}
\\
&\phantom{\mathrm{s.t.}\quad}\mathrm{C4:}~
\sum_{p\in\mathcal{P}_{u}^{\rm R}(n;b(\tau))}
P_{p}^{\rm sky}(n)
\leq P_{u}^{\max},
\quad \forall u,n,
\label{eq:latency-c4}
\\
&\phantom{\mathrm{s.t.}\quad}\mathrm{C5:}~
\mathcal{R}_{k,q}(n)\geq\mathcal{R}_{q}^{\min},
\quad \forall k,q,n,
\label{eq:latency-c5}
\\
&\phantom{\mathrm{s.t.}\quad}\mathrm{C6:}~
0\leq\overline{D}_{k,q}(n)\leq D_{q}^{\max},
\quad \forall k,q,n,
\label{eq:latency-c6}
\\
&\phantom{\mathrm{s.t.}\quad}\mathrm{C7:}~
b(\tau)\in\mathcal{B},
\label{eq:latency-c7}
\end{align}
\end{subequations}
\end{Prob}
\noindent where constraints C1 and C2 define the feasible path-level traffic-splitting ratios, $\mathcal{P}_{k}(n;b(\tau))$ contains only the E2E paths available under the selected sky band. C3 and C4 impose the transmit-power budgets of the ground vehicles and UAVs, respectively. C5 guarantees the minimum E2E transmission reliability of each traffic type, whereas C6 ensures that the prescribed latency requirement is satisfied. C7 restricts the sky-link frequency selection to the available candidate FR3 upper-mid-band set.

On the other hand,  minimizing E2E latency may favor high-rate paths with relatively high outage risks.  We therefore formulate a complementary reliability-oriented optimization problem, in which the E2E reliability is maximized while satisfying the prescribed latency requirements.









\begin{Prob}[E2E Reliability Optimization]
\label{Maximum Reliability}
\small
\begin{subequations}
\label{eq:reliability-objective}
\begin{align}
&\max_{\substack{
\{\boldsymbol{\alpha}(n),\,\mathbf{P}(n)\}_{n\in\mathcal{N}_{\tau}},\,
b(\tau)}}
\quad \overline{\mathcal{R}}_{\mathrm{E2E}}(\tau)
\tag{\theparentequation}
\\[2pt]
&\mathrm{s.t.}\quad
\mathrm{C1-C4} \text{ in} \textbf{ Problem~\ref{Minimum Latency}},
\label{eq:reliability-resource-constraints}
\\
&\phantom{\mathrm{s.t.}\quad}\mathrm{C5}^{\prime}:~
0\leq\overline{D}_{k,q}(n)\leq D_{q}^{\max},
\quad \forall k,q,n,
\label{eq:reliability-c5}
\\
&\phantom{\mathrm{s.t.}\quad}\mathrm{C6}^{\prime}:~
b(\tau)\in\mathcal{B}.
\label{eq:reliability-c6}
\end{align}
\end{subequations}
\end{Prob}

Note that Problem 1 and Problem 2 characterize two complementary extreme operating points: Problem 1 pursues the minimum E2E completion latency while guaranteeing a prescribed reliability level, whereas Problem 2 maximizes the E2E transmission reliability subject to the latency requirements. Therefore, the two problems provide the latency-oriented and reliability-oriented reference operating points, respectively.

\section{Agentic AI-Assisted Hierarchical Learning for
Joint Reliability-Latency Optimization}
\label{secV}

\begin{figure}[t]
    \centering
    \includegraphics[
        width=0.5\textwidth,
    ]{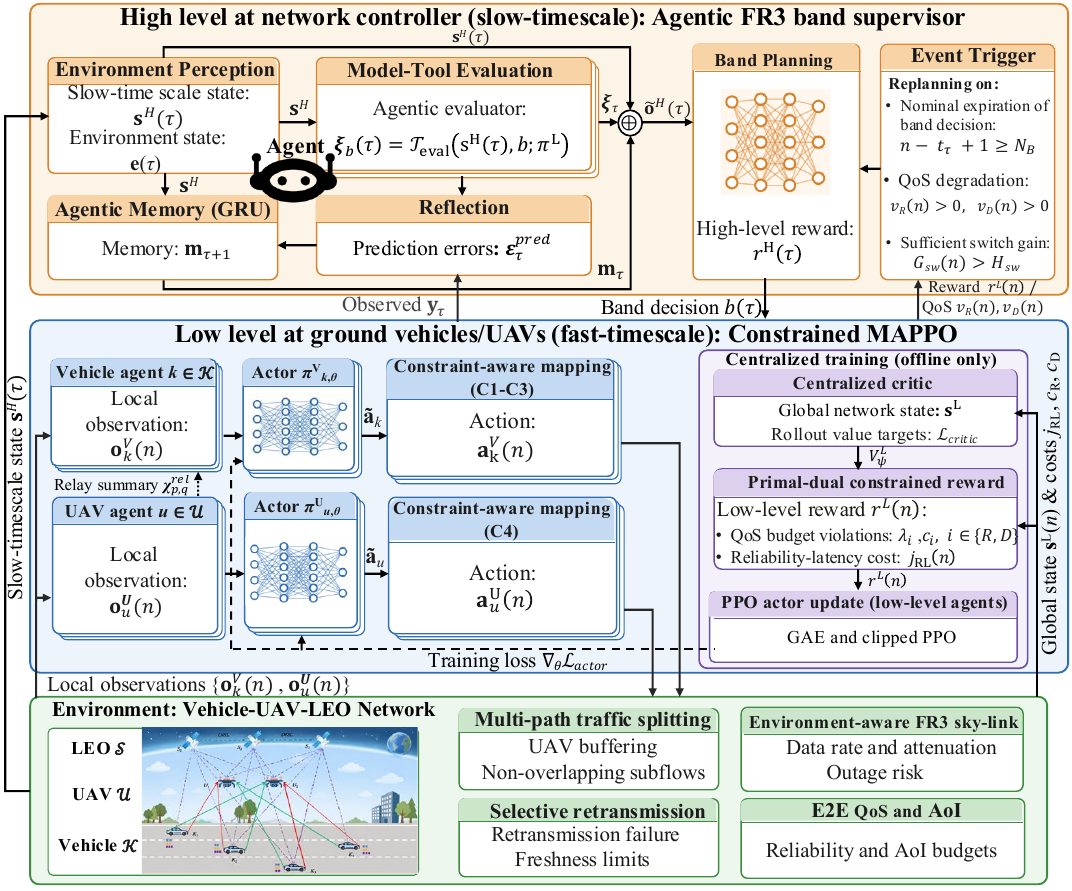}
    \vspace{-0.5cm}
    \caption{The AA-HMARL framework.}
    \label{fig:AAHMARL_arch}
\end{figure}



Although Problems~1 and~2 provide latency- and reliability-oriented reference designs, respectively, practical vehicle-UAV-LEO operations require a dynamic balance between these objectives.  Meanwhile, FR3 band selection follows slow environmental and satellite-availability changes, while traffic splitting and power allocation follow fast channel, queue, and freshness variations. We address this issue through the proposed AA-HMARL.

\subsection{Joint Reliability-Latency Optimization}

To characterize the trade-off between the reference operating points, we define $\phi_{\rm RL}\in[0,1]$ as the reliability-latency preference factor. Define the normalized block-level latency-reliability cost as
\begin{equation}
\small
\begin{aligned}
j_{\rm RL}(n)
=
\frac{1}{K}
\sum_{k\in\mathcal K}
\sum_{q\in\mathcal Q}
\omega_q
\Bigg[
\phi_{\rm RL}
\frac{\bar D_{k,q}(n)}{D_q^{\max}}+
(1-\phi_{\rm RL})
\left(1-\mathcal R_{k,q}(n)\right)
\Bigg].
\end{aligned}
\label{eq:rl-block-cost}
\end{equation}
Then the average cost over the $\tau$-th slow-timescale interval is
\begin{equation}
J_{\rm RL}(\tau)
=
\frac{1}{d_\tau}
\sum_{n\in\mathcal N_\tau}
j_{\rm RL}(n).
\label{eq:rl-interval-cost}
\end{equation}


\begin{remark}
The preference factor $\phi_{\rm RL}$ provides a flexible means of adjusting the operating trade-off. $\phi_{\rm RL}=0$ emphasizes transmission reliability, whereas $\phi_{\rm RL}=1$ emphasizes latency reduction. Increasing $\phi_{\rm RL}$ generally favors shorter E2E completion latency, potentially reducing reliability; decreasing $\phi_{\rm RL}$ will favor reliable delivery, potentially requiring additional transmission or retransmission time.
\end{remark}

Changing the FR3 sky-link band incurs a switching cost
$C_{\rm sw}(b(\tau-1),b(\tau))
=
c_{\rm sw}
\mathbf{1}\{b(\tau)\neq b(\tau-1)\}$,
where $c_{\rm sw}>0$ is the normalized reconfiguration cost. Rather than the hard per-block QoS constraints in Problems~1 and~2, the joint sequential problem uses average violation budgets. Specifically, the block-level reliability violation is
\begin{equation}
c_{\rm R}(n)
=
\frac{1}{K}
\sum_{k\in\mathcal K}
\sum_{q\in\mathcal Q}
\omega_q
\left[
\mathcal R_q^{\min}
-
\mathcal R_{k,q}(n)
\right]^+,
\label{eq:reliability-violation}
\end{equation}
and the block-level normalized latency violation is
\begin{equation}
c_{\rm D}(n)
=
\frac{1}{K}
\sum_{k\in\mathcal K}
\sum_{q\in\mathcal Q}
\omega_q
\left[
\frac{
\bar D_{k,q}(n)-D_q^{\max}
}{
D_q^{\max}
}
\right]^+.
\label{eq:latency-violation}
\end{equation}
Let $\pi=(\pi^{\rm H},\pi^{\rm L})$ denote the hierarchical policy and
$C_\tau\triangleq
C_{\rm sw}(b(\tau-1),b(\tau))$.
As the selected band lasts for a variable duration $d_\tau$, we define the long-term average cost over fast blocks:
\begin{equation}
\bar J_{\rm RL}(\pi)
=
\limsup_{T\to\infty}
\frac{
\mathbb E_\pi
\left[
\sum_{\tau=1}^{T}
\left(
d_\tau J_{\rm RL}(\tau)
+
\lambda_{\rm sw}C_\tau
\right)
\right]
}{
\mathbb E_\pi
\left[
\sum_{\tau=1}^{T}d_\tau
\right]
},
\label{eq:long-term-rl-cost}
\end{equation}
where $\lambda_{\rm sw}\geq 0$ controls the switching-cost penalty, and $T$ is the number of completed high-level decisions. The corresponding average QoS violations of reliability and latency are given by
\begin{equation}
\bar c_i(\pi)
=
\limsup_{T\to\infty}
\frac{
\mathbb E_\pi
\left[
\sum_{\tau=1}^{T}
\sum_{n\in\mathcal N_\tau}
c_i(n)
\right]
}{
\mathbb E_\pi
\left[
\sum_{\tau=1}^{T}d_\tau
\right]
},
\quad
i\in\{\rm R,D\}.
\label{eq:long-term-violation}
\end{equation}
Then the joint optimization problem is formulated as follows.
\begin{Prob}[Joint Reliability-Latency Optimization]
\label{prob:joint-rl}
\small
\begin{subequations}
\label{eq:joint-rl-objective}
\begin{align}
&\min_{\pi^{\rm H},\,\pi^{\rm L}}
\quad \bar J_{\rm RL}(\pi)
\tag{\theparentequation}
\\[2pt]
&\mathrm{s.t.}\quad
\mathrm{C1-C4} \text{ and } \mathrm{C7} \text{ in }\textbf{Problem~\ref{Minimum Latency}},
\label{eq:joint-rl-resource-constraints}
\\
&\phantom{\mathrm{s.t.}\quad}\mathrm{C5}^{''}:
\bar c_{\rm R}(\pi)\leq\epsilon_{\rm R},
\label{eq:joint-rl-c5-prime}
\\
&\phantom{\mathrm{s.t.}\quad}\mathrm{C6}^{''}:
\bar c_{\rm D}(\pi)\leq\epsilon_{\rm D},
\label{eq:joint-rl-c6-prime}
\end{align}
\end{subequations}
\end{Prob}
\noindent where $\mathrm{C5}^{''}$ and  $\mathrm{C6}^{''}$allow average violations $\epsilon_{\rm R}\geq 0$ and $\epsilon_{\rm D}\geq 0$. Note that although the AoI is not an additional objective in Problem~3, it remains in the system dynamics through the AoI-constrained effective retransmission limit in Sec.~III. Information freshness therefore affects both feasible retransmissions and future reliability/latency performance without introducing an additional objective.

\subsection{Agentic Mid-band Supervision}
The supervisor selects the slow-timescale sky-link band $b(\tau)$; conditioned on this band, vehicle/UAV agents select $\boldsymbol{\alpha}(n)$ and $\mathcal{P}(n)$ in each block. Let $\pi^{\rm H}$ be the supervisor policy and $\boldsymbol{\pi}^{\rm L}
    =
    \left\{
    \pi_k^{\rm V}
    \right\}_{k\in\mathcal{K}}
    \cup
    \left\{
    \pi_u^{\rm U}
    \right\}_{u\in\mathcal{U}}$
denote the low-level ground vehicle/UAV policies. Conditioned on the current band $b(\tau)$, the joint low-level policy is factorized as
\begin{equation}
\small
\begin{split}
&
\Pr
\Big(
b(\tau),
\{
\mathbf{a}^{\rm L}(n)
\}_{n\in\mathcal{N}_{\tau}}
\Big)
\\
&=
\pi^{\rm H}
\left(
b(\tau)
\mid
\widetilde{\mathbf{o}}^{\rm H}(\tau)
\right)
\prod_{n\in\mathcal{N}_{\tau}}
\left[
\prod_{k\in\mathcal{K}}
\pi_k^{\rm V}
\left(
\mathbf{a}_k^{\rm V}(n)
\mid
\mathbf{o}_k^{\rm V}(n),b(\tau)
\right)
\right.
\\
&\hspace{35mm}
\left.
\times
\prod_{u\in\mathcal{U}}
\pi_u^{\rm U}
\left(
\mathbf{a}_u^{\rm U}(n)
\mid
\mathbf{o}_u^{\rm U}(n),b(\tau)
\right)
\right].
\end{split}
\label{eq:hierarchical_policy}
\end{equation}

The supervisor combines environment perception, long-term memory, model-tool evaluation, and event-triggered replanning to exploit learned experience and the analytical models of Sec.~III. At the start of interval $\tau$, it observes the slow-timescale state:
\begin{equation}
\begin{split}
    \mathbf{s}^{\rm H}(\tau)
    =
    \Big[
    &\mathbf{e}(\tau),
    b(\tau-1),
    \{\overline{p}_{p,b,q}\},
    \\
    &\{\overline{D}_{k,q}\},
    \{\mathcal{R}_{k,q}\},
    \{\Delta_{k,q}\},
    \mathbf{m}^{\rm path}(\tau)
    \Big],
\end{split}
\label{eq:high_state}
\end{equation}
where $\mathbf{e}(\tau)=[R(\tau),I_{\rm eff}(\tau)]$ is the environmental state, $\mathbf{m}^{\rm path}(\tau)$ masks satellite/path availability, and the remaining entries describe predicted link risks, E2E latency/reliability performance, and information freshness. A recurrent memory vector $\mathbf{m}_{\tau}$ retains historical relations between environment and action performance. After interval $\tau$, the observed outcome summary is
\begin{equation}
    \mathbf{y}_{\tau}
    =
    \left[
    J_{\rm RL}(\tau),
    \overline{D}_{\rm E2E}(\tau),
    \overline{\mathcal{R}}_{\rm E2E}(\tau),
    \overline{\Delta}(\tau)
    \right],
\label{eq:supervisor_outcome}
\end{equation}
and the memory is updated through a gated recurrent unit (GRU) $f_{\rm mem}(\cdot)$ as
\begin{equation}
\mathbf m_{\tau+1}
=
f_{\rm mem}
\left(
\mathbf m_\tau,
\mathbf s^{\rm H}(\tau),
b(\tau),
\mathbf y_\tau,
\boldsymbol\varepsilon_\tau^{\rm pred}
\right),
\label{eq:memory_update}
\end{equation}
where $\boldsymbol\varepsilon_\tau^{\rm pred}$ is the prediction-error vector. For each candidate FR3 band $b\in\mathcal{B}$, the channel, risk, retransmission, and latency models in Sec.~III provide an agentic evaluator
\begin{equation}
\begin{split}
    \boldsymbol{\xi}_{b}(\tau)
    &=
    \mathcal{T}_{\rm eval}
    \left(
    \mathbf{s}^{\rm H}(\tau),
    b;
    \boldsymbol{\pi}^{\rm L}
    \right)
    =
    \left[
    \widehat{D}_{b}(\tau),
    \widehat{\mathcal{R}}_{b}(\tau),
    \widehat{J}_{b}(\tau)
    \right],
\end{split}
\label{eq:model_tool}
\end{equation}
where $\widehat{D}_{b}$, $\widehat{\mathcal{R}}_{b}$, and $\widehat{J}_{b}$ are predicted E2E completion latency, reliability, and joint cost, respectively. Short-horizon analytical rollouts with the current low-level policy can implement the evaluator. After a high-level action, selected-band predictions are compared with observed outcomes by
\begin{equation}
\boldsymbol\varepsilon_\tau^{\rm pred}
=
\begin{bmatrix}
\bar D_{\rm E2E}(\tau)
-\widehat D_{b(\tau)}(\tau)\\
\bar{\mathcal R}_{\rm E2E}(\tau)
-\widehat{\mathcal R}_{b(\tau)}(\tau)\\
J_{\rm RL}(\tau)
-\widehat J_{b(\tau)}(\tau)
\end{bmatrix},
\label{eq:prediction_error}
\end{equation}
which are incorporated into the memory update in Eq. \eqref{eq:memory_update}, allowing the supervisor to learn from discrepancies between model-based estimates and actual outcomes. The tool outputs of all candidate bands are collected as $\boldsymbol{\xi}_{\tau}
    =
    \left\{
    \boldsymbol{\xi}_{b}(\tau)
    \right\}_{b\in\mathcal{B}}$
, which gives the augmented observation $\widetilde{\mathbf{o}}^{\rm H}(\tau)
    =
    \left[
    \mathbf{s}^{\rm H}(\tau),
    \mathbf{m}_{\tau},
    \boldsymbol{\xi}_{\tau}
    \right]$.
This observation combines the current environment, historical experience, and model-based candidate-band evaluation for high-level decisions.
The supervisor selects
$b(\tau)
    \sim
    \pi^{\rm H}
    \left(
    \cdot
    \mid
    \widetilde{\mathbf{o}}^{\rm H}(\tau);
    \boldsymbol{\theta}_{\rm H}
    \right)$,
with policy parameters $\boldsymbol{\theta}_{\rm H}$ and an action mask excluding bands without feasible sky paths.

A band action normally lasts $N_{\rm B}$ fast blocks but can terminate earlier once the reliability/latency degrades significantly. The reliability and normalized latency violations used to detect such degradation are expressed as
\begin{equation}
    v_{\rm R}(n)
    =
    \max_{k,q}
    \left[
    \mathcal{R}_{q}^{\min}
    -
    \mathcal{R}_{k,q}(n)
    \right]^+,
\label{eq:rel_violation}
\end{equation}
\begin{equation}
    v_{\rm D}(n)
    =
    \max_{k,q}
    \left[
    \frac{
    \overline{D}_{k,q}(n)
    -
    D_q^{\max}
    }{
    D_q^{\max}
    }
    \right]^+.
\label{eq:lat_violation}
\end{equation}
The predicted replanning benefit under the current environmental and QoS conditions is
\begin{equation}
\begin{split}
    G_{\rm sw}(n)
    =
    \widehat{J}_{b_{\rm cur}}(n)
    -
    \min_{b\in\mathcal{B}}
    \Big[
    \widehat{J}_{b}(n)
    +
    \lambda_{\rm sw}c_{\rm sw}
    \mathbf{1}
    \{b\neq b_{\rm cur}\}
    \Big],
\end{split}
\label{eq:switch_gain}
\end{equation}
and the termination indicator is
\begin{equation}
\begin{split}
    \beta_{\tau}(n) = \mathbf{1} \Big\{
    &(n-t_{\tau}+1)\geq N_{\rm B}
    \lor\; v_{\rm R}(n)>0\\
    &
    \lor\; v_{\rm D}(n)>0\lor\; G_{\rm sw}(n)>H_{\rm sw} \Big\},
\end{split}
\label{eq:option_termination}
\end{equation}
where $H_{\rm sw}>0$ is the hysteresis threshold. The option expires after $N_{\rm B}$ blocks or terminates early for QoS violations or sufficient switching gain. When QoS remains satisfied, hysteresis suppresses switches driven by small predicted gains. Accordingly, if the current high-level action starts at block $t_\tau$ and terminates at block $n_\tau^{\rm end}$, its actual duration is
$d_\tau
=
n_\tau^{\rm end}-t_\tau+1$. Thus the slow-timescale control process is modeled as a semi-Markov decision process (SMDP)\cite{SUTTON1999181}.

\subsection{Feasible Multi-Agent Resource Allocation}
Conditioned on $b(\tau)$, ground vehicle and UAV agents jointly select traffic splitting ratios and transmit powers each fast block. Ground vehicle $k$ combines local V2U/V2S measurements with relay-path summaries reported by associated UAVs. This supports splitting across direct and relay paths while keeping detailed UAV states and power decisions local. For $p=(k,u,s)$, the summary is
\begin{equation}
\boldsymbol{\chi}_{p,q}^{\rm rel}(n)
=
\left[
\widehat R_{p,b}^{\rm sky}(n),
\widehat{\bar p}_{p,b,q}(n),
\widehat d_{p,q}^{\rm que}(n)
\right],
\label{eq:relay-quality-summary}
\end{equation}
where the entries predict the U2S rate, conservative outage-risk envelope, and UAV queueing delay using the channel, interference, and buffer information available before the current decisions. The observation of ground vehicle $k$ is expressed as
\begin{equation}
\small
\begin{aligned}
&\mathbf{o}_k^{\rm V}(n)
=
\Big[
\{\gamma_{k,u}^{\rm V2U}(n)\},\{\widehat\gamma_{p,b}^{\rm sky}(n)\}
,\{\widehat{\bar p}_{p,b,q}(n)\}
,\\
&\{\boldsymbol{\chi}_{p,q}^{\rm rel}(n)\}
,\{\Xi_{k,q}(n),\Delta_{k,q}(n)\},\{z_{p,q}(n),r_{p,q}(n)\},
b(\tau)
\Big],
\end{aligned}
\label{eq:vehicle-observation}
\end{equation}
where $\widehat{\gamma}_{p,b(n)}^{\mathrm{sky}}(n)$ denotes the estimated sky-link SINR available to the vehicle agent before the current resource allocation.

Meanwhile, the action of ground vehicle $k$ is expressed as
\begin{equation}
    \mathbf{a}_{k}^{\rm V}(n)
    =
    \left[
    \boldsymbol{\alpha}_{k}(n),
    \{P_{k,u}^{\rm V2U}(n)\},
\{P_{k,s}^{\rm V2S}(n)\}
    \right].
\label{eq:vehicle_action}
\end{equation}
For new packets, the ground vehicle determines $\boldsymbol\alpha_k(n)$; selective retransmissions retain the original path assignment and traffic ratios, updating only the corresponding powers. UAV agents control forwarding of unresolved buffered subflows, without repeating V2U transmission after a sky-hop failure. UAV $u$ observes its buffered-subflow, sky-link, retransmission, and AoI states:
\begin{equation}
\begin{split}
    \mathbf{o}_{u}^{\rm U}(n)
    =
    \Big[
    &\{B_{p,q}(n)\},
    \{\gamma_{p,b}^{\rm sky}(n)\},
    \{\overline{p}_{p,b,q}(n)\},\\
    &
    \{z_{p,q}(n)\}, \{r_{p,q}(n)\},
    \{\Delta_{k,q}(n)\},
    b(\tau)
    \Big],
\end{split}
\label{eq:uav_obs}
\end{equation}
and periodically reports the summaries in Eq. \eqref{eq:relay-quality-summary}, retaining detailed local states and power allocation decisions. The action of UAV $u$ is given by
\begin{equation}
    \mathbf{a}_{u}^{\rm U}(n)
    =
    \left[\{
    P_{k,u,s}^{\rm U2S}(n)
    \}\right].
\label{eq:uav_action}
\end{equation}

During centralized training, the critic observes:
\begin{equation}
\begin{split}
    \mathbf{s}^{\rm L}(n)
    =
    \Big[
    &\mathbf{e}(\tau),
    b(\tau),
    \mathbf{H}^{\rm V2U}(n),
    \boldsymbol{\Gamma}^{\rm sky}(n),
    \\
    &\overline{\mathbf{p}}(n),
    \boldsymbol{\Xi}(n),
    \mathbf B(n),
    \boldsymbol{\Delta}(n),
    \mathbf{z}(n),
    \mathbf{r}(n)
    \Big],
\end{split}
\label{eq:global_state}
\end{equation}
where $\boldsymbol\Gamma^{\rm sky}$, $\bar{\mathbf p}$, $\mathbf z$, and $\mathbf r$ collect sky-link SINRs, outage-risk envelopes, unresolved indicators, and retransmission states.  

To reduce infeasible actions during learning, the constraints C1-C4 in Problem~1 are satisfied directly through the action parameterization. Let $x_{p,q}(n)$ denote the raw traffic-splitting logit generated for path $p\in\mathcal P_k(n)$, and $m_p(n;b)\in\{0,1\}$ denote the corresponding feasible path mask. The actual traffic-splitting ratios are obtained through a masked sparsemax mapping: 
\begin{equation}
\alpha_{p,q}(n)
=
\begin{cases}
\left[x_{p,q}(n)-\theta_{k,q}(n)\right]^+,
& m_p(n;b)=1,\\
0, & m_p(n;b)=0,
\end{cases}
\label{eq:alpha-projection}
\end{equation}
where the threshold $\theta_{k,q}(n)$ uniquely enforces $\sum_{p\in\mathcal P_k(n;b)}\alpha_{p,q}(n)=1$. Thus the constraints C1-C2 hold, while exactly zero ratios permit sparse active-path selection. Unlike softmax, sparsemax can suppress unfavorable feasible paths entirely, allowing the active set to be a strict subset of the available paths\cite{pmlr-v48-martins16}. Let $\mathcal L_k^{\rm out}(n)$ contain the vehicle's active V2U/V2S links and $\widetilde P_{k,l}(n)$ denote raw power logits. The vehicle mapping is
\begin{equation}
\small
    P_{k,l}(n)
    =
    P_{k}^{\max}
    \frac{
    \exp
    \left(
    \widetilde{P}_{k,l}(n)
    \right)
    }{
    1+
    \displaystyle
    \sum_{l'\in\mathcal{L}_{k}^{\rm out}(n)}
    \exp
    \left(
    \widetilde{P}_{k,l'}(n)
    \right)
    }.
\label{eq:vehicle_power_projection}
\end{equation}
As a result, $\sum_{l\in\mathcal{L}_{k}^{\rm out}(n)}
    P_{k,l}(n)
    \leq
    P_k^{\max}$. For UAV $u$, the corresponding mapping is given by
\begin{equation}
\small
    P_{p}^{\rm sky}(n)
    =
    P_{u}^{\max}
    \frac{
    \exp
    \left(
    \widetilde{P}_{p}(n)
    \right)
    }{
    1+
    \displaystyle
    \sum_{p'}
    \exp
    \left(
    \widetilde{P}_{p'}(n)
    \right)
    },\quad p \in \mathcal{P}_{u}^{R}(n),
\label{eq:uav_power_projection}
\end{equation}
where the denominator sums over UAV $u$'s scheduled relay paths, thus its outgoing powers share the same budget and satisfy the constraint C4 in Problem 1. Resource feasibility is therefore enforced by the actor outputs rather than only reward penalties. The high-level band mask handles the constraint C7, while the long-term QoS constraints $\mathrm{C5}^{''}$-$\mathrm{C6}^{''}$ are addressed approximately through primal-dual learning. For agent $i$, the stochastic policy $\pi_{i,\theta}^{\rm L}$ samples a raw latent action $\widetilde{\mathbf a}_i(n)$, mapped to physical resources by $\mathbf a_i(n)
=
g_i\left(\widetilde{\mathbf a}_i(n)\right)$,
where $g_i(\cdot)$ implements the traffic-splitting and power allocation mappings.

\subsection{Hierarchical Policy Learning}
\subsubsection{Low-Level Constrained MAPPO}
The Lagrangian reward combines the block cost and QoS violations\cite{NEURIPS2024_fa76985f} defined in Sec.~V-A:
\begin{equation}
r^{\rm L}(n)
=
-j_{\rm RL}(n)
-\lambda_{\rm R}c_{\rm R}(n)
-\lambda_{\rm D}c_{\rm D}(n),
\label{eq:low-level-reward}
\end{equation}
where the nonnegative multipliers $\lambda_{\rm R},\lambda_{\rm D}$ adapt to observed violations. Their projected updates are
\begin{equation}
\lambda_{\rm R}^{(t+1)}
=
\left[
\lambda_{\rm R}^{(t)}
+
\eta_{\rm R}
\left(
\widehat c_{\rm R}^{(t)}
-
\epsilon_{\rm R}
\right)
\right]^+,
\label{eq:lambda-r-update}
\end{equation}
\begin{equation}
\lambda_{\rm D}^{(t+1)}
=
\left[
\lambda_{\rm D}^{(t)}
+
\eta_{\rm D}
\left(
\widehat c_{\rm D}^{(t)}
-
\epsilon_{\rm D}
\right)
\right]^+,
\label{eq:lambda-d-update}
\end{equation}
where $\eta_{\rm R},\eta_{\rm D}$ are dual learning rates, and $\widehat c_{\rm R}^{(t)},\widehat c_{\rm D}^{(t)}$ are empirical average violations from training rollouts. The updates respond to violations relative to their budgets while projection maintains nonnegativity. With centralized training and decentralized execution, the band-conditioned joint policy factorizes as
\begin{equation}
\begin{split}
    \pi^{\rm L}
    \left(
    \mathbf{a}^{\rm L}(n)
    \mid
    \mathbf{s}^{\rm L}(n),b(\tau)
    \right)
    =
    &
    \prod_{k\in\mathcal{K}}
    \pi_k^{\rm V}
    \left(
    \mathbf{a}_k^{\rm V}(n)
    \mid
    \mathbf{o}_k^{\rm V}(n),b(\tau)
    \right)
    \\
    &\times
    \prod_{u\in\mathcal{U}}
    \pi_u^{\rm U}
    \left(
    \mathbf{a}_u^{\rm U}(n)
    \mid
    \mathbf{o}_u^{\rm U}(n),b(\tau)
    \right).
\end{split}
\label{eq:low_joint_policy}
\end{equation}

The critic exploits the global channels, queue, freshness, and retransmission information during training, while the actor's inference remains local. For the centralized value function $V_\psi^{\rm L}(\mathbf s^{\rm L}(n))$, the temporal-difference error is
\begin{equation}
\begin{split}
    \delta^{\rm L}(n)
    =
    r^{\rm L}(n)
    +
    \gamma_{\rm L}
    V_{\psi}^{\rm L}
    \left(
    \mathbf{s}^{\rm L}(n+1)
    \right)
    -
    V_{\psi}^{\rm L}
    \left(
    \mathbf{s}^{\rm L}(n)
    \right).
\end{split}
\label{eq:low_td}
\end{equation}
The generalized advantage estimate (GAE) is given by
\begin{equation}
    \widehat{A}^{\rm L}(n)
    =
    \sum_{l=0}^{\infty}
    \left(
    \gamma_{\rm L}\lambda_{\rm GAE}
    \right)^l
    \delta^{\rm L}(n+l).
\label{eq:low_gae}
\end{equation}

As $g_i$ can be non-invertible, PPO evaluates probabilities using sampled raw latent actions:
\begin{equation}
\rho_i^{\rm L}(n)
=
\frac{
\pi_{i,\theta}^{\rm L}
\left(
\widetilde{\mathbf a}_i(n)
\mid \mathbf o_i(n)
\right)
}{
\pi_{i,\theta_{\rm old}}^{\rm L}
\left(
\widetilde{\mathbf a}_i(n)
\mid \mathbf o_i(n)
\right)
}.
\label{eq:ppo-ratio}
\end{equation}
Rollouts store these actions and their log-probabilities while the environment executes the mapping actions. The probability ratio compares the current latent policy with the policy that was generated by the rollout, avoiding inversion of the resource mapping. The clipped actor objective is
\begin{equation}
\begin{split}
    \mathcal{L}_{i}^{\rm actor}
    =
    -
    \mathbb{E}
    \Big[
    \min
    \big(
    \rho_i^{\rm L}
    \widehat{A}^{\rm L},
    \operatorname{clip}
    (
    \rho_i^{\rm L},
    1-\epsilon_{\rm PPO},
    1+\epsilon_{\rm PPO}
    )
    \widehat{A}^{\rm L}
    \big)
    \Big],
\end{split}
\label{eq:low_actor_loss}
\end{equation}
and the centralized critic minimizes
\begin{equation}
    \mathcal{L}_{\rm critic}^{\rm L}
    =
    \mathbb{E}
    \left[
    \left(
    V_{\psi}^{\rm L}
    \left(
    \mathbf{s}^{\rm L}(n)
    \right)
    -
    \widehat{G}^{\rm L}(n)
    \right)^2
    \right],
\label{eq:low_critic_loss}
\end{equation}
where $\widehat G^{\rm L}(n)$ is the return target.

\subsubsection{High-Level SMDP Learning}
For an option of actual duration $d_\tau$, the high-level reward accumulates discounted low-level rewards and the switching cost:
\begin{equation}
\begin{split}
    r^{\rm H}(\tau)
    =
    \sum_{j=0}^{d_{\tau}-1}
    \gamma_{\rm L}^{j}
    r^{\rm L}(t_{\tau}+j)
    -
    \lambda_{\rm sw}
    C_{\rm sw}
    \big(
    b(\tau-1),b(\tau)
    \big).
\end{split}
\label{eq:high_reward}
\end{equation}
Then the agentic supervisor is trained to maximize the
long-term return $\max_{\pi^{\rm H}}
    \;
    \mathbb{E}_{\pi^{\rm H}}
    \left[
    \sum_{\tau=0}^{\infty}
    \gamma_{\rm H}^{t_{\tau}}
    r^{\rm H}(\tau)
    \right]$. For the high-level value function $V_{\psi_{\rm H}}^{\rm H}(\widetilde{\mathbf o}^{\rm H}(\tau))$, the SMDP temporal-difference error accounts for the variable option duration:
\begin{equation}
\begin{split}
    \delta^{\rm H}(\tau)
    =
    r^{\rm H}(\tau)
    +
    \gamma_{\rm H}^{d_{\tau}}
    V_{\psi_{\rm H}}^{\rm H}
    \left(
    \widetilde{\mathbf{o}}^{\rm H}(\tau+1)
    \right)
    -
    V_{\psi_{\rm H}}^{\rm H}
    \left(
    \widetilde{\mathbf{o}}^{\rm H}(\tau)
    \right).
\end{split}
\label{eq:high_td}
\end{equation}
The corresponding high-level advantage is obtained by SMDP-aware GAE, and the discrete band selection policy $\pi^{\rm H}$ is updated using the PPO clipped surrogate objective in the same manner as
Eq. \eqref{eq:low_actor_loss}. After each high-level action terminates, the supervisor performs a reflection step by comparing the predicted tool outputs $\{\widehat{D}_{b}, \widehat{\mathcal{R}}_{b}, \widehat{J}_{b}\}$ with the actually observed outcome $\mathbf{y}_{\tau}$, and then updates the memory according to Eq. \eqref{eq:memory_update}. In this way, prediction errors and historical switching outcomes are retained in the memory and can influence subsequent band-planning decisions.


\begin{algorithm}[t]
\caption{Agentic AI-Assisted Hierarchical MARL}
\label{alg:aahmarl}
\footnotesize
\begin{algorithmic}[1]
\Require Candidate bands $\mathcal B$, $\phi_{\rm RL}$, $N_{\rm B}$, $H_{\rm sw}$,
rollout lengths $L_{\rm L},L_{\rm H}$, and PPO epochs $E_{\rm L},E_{\rm H}$
\Ensure Trained policies $\pi^{\rm H}$ and $\boldsymbol{\pi}^{\rm L}$

\State Initialize policies, critics, $\mathbf m_0$, and $\lambda_{\rm R},\lambda_{\rm D}$
\Statex \textbf{Stage 1: Low-level pre-training}
\For{each low-level training rollout}
    \State Initialize the low-level buffer
    \While{fewer than $L_{\rm L}$ transitions are collected}
        \If{a new episode is required}
            \State Reset environment and sample a candidate band
        \EndIf
        \State Obtain vehicle/UAV observations
        \State Sample raw actions and store actions and log-probabilities
        \State Map actions and execute V2U/V2S/U2S transmissions
        \State Perform selective retransmissions
        \State Compute $r^{\rm L}(n)$ and store the transition
    \EndWhile
    \State Compute low-level GAE with episode masks
    \State Update MAPPO actors and centralized critic for $E_{\rm L}$ epochs
    \State Update $\lambda_{\rm R},\lambda_{\rm D}$ and clear the buffer
\EndFor

\Statex \textbf{Stage 2: Agentic supervisor training}
\State Fix $\boldsymbol{\pi}^{\rm L}$; initialize the high-level buffer
\For{each high-level training episode}
    \State Reset environment and agentic memory
    \While{the episode is not terminated}
        \State Observe $\mathbf s^{\rm H}(\tau)$
        \State Evaluate candidate bands using $\mathcal T_{\rm eval}$
        \State Form $\widetilde{\mathbf o}^{\rm H}(\tau)$; sample $b(\tau)$ from masked $\pi^{\rm H}$
        \State Store band predictions and log-probability
        \Repeat
            \State Run $\boldsymbol{\pi}^{\rm L}$ with $b(\tau)$; log rewards, QoS, and AoI
        \Until{$\beta_\tau(n)=1$ or episode termination}
        \State Compute $d_\tau$, $r^{\rm H}(\tau)$, and $\boldsymbol\varepsilon_\tau^{\rm pred}$
        \State Update $\mathbf m_{\tau+1}$; store the option transition
        \If{the high-level buffer reaches $L_{\rm H}$ options}
            \State Compute high-level GAE
            \State Update high-level PPO actor and critic for $E_{\rm H}$ epochs
            \State Clear the high-level buffer
        \EndIf
    \EndWhile
\EndFor
\State Update high-level PPO using any remaining rollout data

\Statex \textbf{Stage 3: Joint fine-tuning}
\While{the training criterion is not satisfied}
    \State Collect hierarchical interaction rollouts
    \State Update memory after completed high-level options
    \State Compute low-level and SMDP-aware high-level GAE
    \State Update low-level MAPPO actors and critic for $E_{\rm L}$ epochs
    \State Update high-level PPO actor and critic for $E_{\rm H}$ epochs
    \State Update $\lambda_{\rm R},\lambda_{\rm D}$
\EndWhile
\State \Return $\pi^{\rm H}$ and $\boldsymbol{\pi}^{\rm L}$
\end{algorithmic}
\end{algorithm}

\subsection{Training Procedure and Computational Complexity}
Directly training both hierarchical levels from scratch may introduce severe non-stationarity because the operating environment observed by the high-level supervisor changes while the low-level policies are still evolving. We therefore adopt a staged training procedure. First, low-level actors are pretrained under randomly sampled bands and environmental states. Second, they are fixed or updated more slowly while the supervisor learns from the augmented observation $\widetilde{\mathbf o}^{\rm H}(\tau)$. This stabilizes the supervisor's learning environment while preserving the low-level behavior acquired under different channel conditions. Finally, both levels are fine-tuned through alternating policy and dual updates. Low-level training uses fast-block rollouts with episode masks for GAE; meanwhile, high-level training uses completed band-option transitions. At both timescales, PPO updates follow batch collection rather than individual blocks or options, as summarized in Algorithm~\ref{alg:aahmarl}. During online execution, the supervisor evaluates bands only at interval starts or when Eq. \eqref{eq:option_termination} triggers re-evaluation; meanwhile, vehicle/UAV actors execute locally each fast block. Let $N_{\rm A}=K+U$ and $C_{\rm L}$, $C_{\rm H}$, and $C_{\rm T}$ denote the costs of one low-level actor, the high-level actor, and one candidate-band tool evaluation. Normal fast-block inference costs $\mathcal{O}
    \left(
    N_{\rm A}C_{\rm L}
    \right)$,
while replanning adds
$\mathcal{O}
    \left(
    C_{\rm H}
    +
    |\mathcal{B}|C_{\rm T}
    \right)$.
Infrequent supervisor activation reduces the average model-tool evaluation overhead across fast blocks.

\section{Simulation Results}
\label{secVI}

\begin{table}[!t]
\centering
\caption{Main Simulation Parameters}
\vspace{-0.2cm}
\label{tab:sim_parameters}
\tiny
\setlength{\tabcolsep}{3pt}
\renewcommand{\arraystretch}{1.08}
\begin{tabular}{|
>{\centering\arraybackslash}m{0.55\columnwidth}|
>{\centering\arraybackslash}m{0.37\columnwidth}|}
\hline
\textbf{Parameter} & \textbf{Value} \\
\hline

Number of vehicles/UAVs/LEOs, $K/U/S$
& 4/2/3 \\ \hline

Vehicle/UAV/LEO altitude
& 1.5 m / 100 m / 600 km \\ \hline

Vehicle ULA / UAV UPA antennas
& 4 / 4$\times$4 \\ \hline

V2U carrier frequency over Sidelink band, $f_{\rm A}$
& 5.9 GHz \\ \hline

V2U bandwidth, $W_{\rm A}$
& 10 MHz \\ \hline

Candidate FR3 frequencies, $f_b$ (GHz)
& \{7.125,8.5,10.2,13,15.7,18,21,24.25\} \\ \hline

FR3 bandwidths, $W_b$ (MHz)
& \{20,20,30,40,50,60,70,80\} \\ \hline

Vehicle/UAV power budget,
$P_k^{\max}/P_u^{\max}$
& 0.2/2 W \\ \hline

Noise PSD, $N_0$
& -174 dBm/Hz \\ \hline

V2U path-loss exponent, $\eta_{\rm V2U}$
& 2.2 \\ \hline

Shadow-fading standard deviation
& 4 dB \\ \hline

Rain intensity (mm/h)
& \{0,10,25\} \\ \hline

V2X traffic types, $q$
& \{1,2\} \\ \hline

Packet size, $\Xi_{k,q}$ (bytes)
& 1500/6000 \\ \hline

Update generation period (ms)
& 100/200 \\ \hline

E2E latency budget, $D_q^{\max}$ (ms)
& 120/200 \\ \hline

AoI deadline, $\Delta_{k,q}^{\max}$ (ms)
& 200/500 \\ \hline

Maximum retransmission rounds,
$R_{p,q}^{\max}$
& 4 \\ \hline

Fast-block duration (ms) / slots per block
& 10/4 \\ \hline

Nominal band holding duration, $N_B$
& 8 blocks \\ \hline

Switching threshold, $H_{\rm sw}$
& 0.05 \\ \hline

Preference factor, $\phi_{\rm RL}$
& [0,1] \\ \hline


Normalized band-switching cost, $c_{\mathrm{sw}}$ & 0.1 \\ \hline

Band-switching penalty weight, $\lambda_{\mathrm{sw}}$ & 0.5 \\ \hline

Agentic supervisor model
& Jev 1.14.0 (TypeSafe AI)\cite{deusser2026evaluating} \\ \hline

Discount factor / PPO clip
& 0.99/0.2 \\ \hline

Actor/critic learning rates
& 3$\times10^{-4}/10^{-3}$ \\ \hline

Rollout lengths, $L_{\rm L}/L_{\rm H}$
& 128/32 \\ \hline

PPO update epochs, $E_{\rm L}/E_{\rm H}$
& 4/4 \\ \hline

Training epochs
& 1800 \\ \hline

Evaluation horizon / Monte Carlo runs
& 60 s / 50 \\ \hline

\end{tabular}
\end{table}



In this section, we evaluate the proposed AA-HMARL framework in the integrated vehicle-UAV-LEO network. We first examine the stability and reliability-latency trade-offs. We then compare E2E completion latency, delivery reliability, and AoI under the three optimization problems. Finally, we investigate the AoI evolution and the dynamic FR3 mid-band selection behavior.

\subsection{Simulation Setup and Baselines}
\label{subsec:simulation_setup}

Table~\ref{tab:sim_parameters} summarizes the network, traffic, channel model, and training parameters used for the simulations. Each vehicle generates two heterogeneous traffic types and may split an update into non-overlapping subflows over direct V2S and UAV-assisted V2U-U2S paths. The V2U band is fixed, while sky links use candidate FR3 bands. Rates and outage estimates follow the frequency-dependent channel and interference models in Sec.~III. The agentic supervisor incorporates the pretrained Jev decision model to assess candidate FR3 bands using environmental states and
analytical performance estimates, with its outputs augmenting
the observation of the PPO-based band-selection policy. Each problem formulation is evaluated with four schemes:
\begin{itemize}
\item \emph{Single-Path (SP):} A fixed feasible E2E path carries each update without splitting.
\item \emph{Path Selection (PS):} One feasible E2E path is adaptively selected using estimated transmission quality, without splitting an update across paths.
\item \emph{Fixed-Band MAPPO (FB-MAPPO):} MAPPO learns multi-path splitting and power allocation with a fixed FR3 band. No high-level supervisor is used.
\item \emph{AA-HMARL:} The full controller combines slow-timescale agentic FR3 selection and fast-timescale multi-agent splitting/power control with memory, model tools, and event-triggered replanning.
\end{itemize}

Problems~1 and~2 target latency and reliability, respectively, with hard QoS requirements; Problem~3 uses a joint cost, average QoS-violation budgets, and a switching penalty. Comparisons \emph{within} each problem isolate transmission strategy effects; comparisons \emph{across} problems illustrate different objectives and constraint regimes. Unless stated otherwise, all schemes use identical traffic-generation processes, network realizations, and environmental traces.

\subsection{Reliability-Latency Tradeoff and Learning Performance}
\label{subsec:sim_convergence}
\label{subsec:sim_phi_tradeoff}

\begin{figure}[t]
    \centering
    \includegraphics[width=0.28\textwidth]{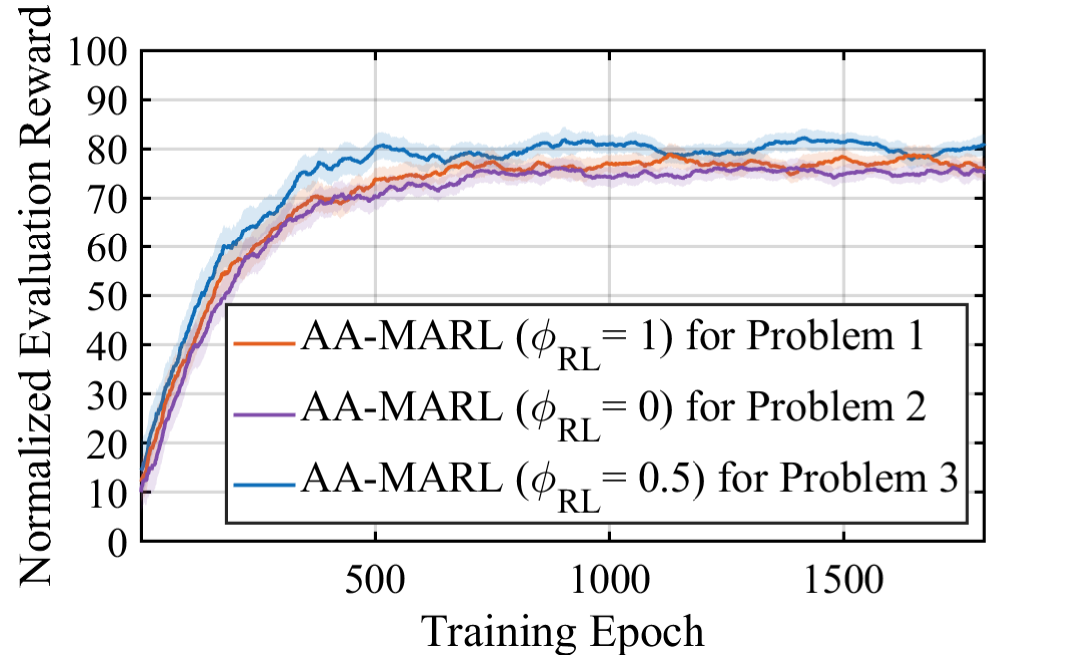}
        \vspace{-0.2cm}
    \caption{Convergence of the normalized evaluation reward.} \label{fig:reward_convergence}
    
\end{figure}

Fig.~\ref{fig:reward_convergence} compares AA-HMARL training with $\phi_{\rm RL}=0$, $0.5$, and $1$, representing reliability-oriented, balanced, and latency-oriented preferences, respectively. Because their training objectives differ, all policies are evaluated using the common normalized reward $R_{\rm eval}
=100\left(1-\frac{J_{\rm eval}}{J_{\rm ref}}\right)$,
where $J_{\rm eval}$ uses identical test scenarios, a fixed evaluation preference factor, and fixed QoS-violation penalties; $J_{\rm ref}>0$ is a common reference cost. The rewards rise rapidly and stabilize with small fluctuations, indicating empirical training stability. The balanced configuration ($\phi_{\rm RL}=0.5$) reaches the highest stable reward under this common evaluation criterion; the extreme preferences reach slightly lower levels.


\begin{figure}[!t]
    \centering
    \includegraphics[width=0.6\columnwidth]{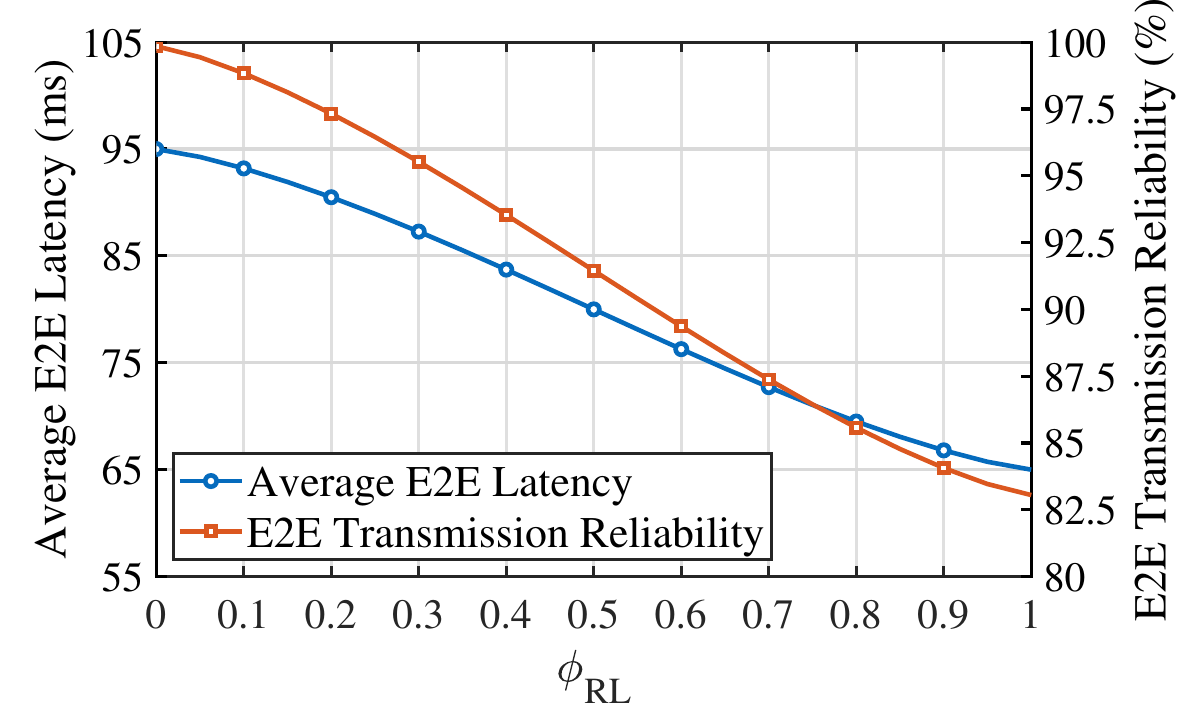}
    \vspace{-0.2cm}
    \caption{Impact of $\phi_{\rm RL}$ on latency and reliability performance.}
    \label{fig:phi_tradeoff}
\end{figure}

To examine the trade-off, we vary $\phi_{\rm RL}$ from $0$ to $1$ with network configurations and QoS constraints unchanged, measuring average E2E completion latency and transmission reliability.
In Fig.~\ref{fig:phi_tradeoff}, increasing $\phi_{\rm RL}$ decreases both latency and reliability, showing the cost of aggressive latency-oriented operation. Smaller values favor reliable transmission, potentially requiring longer transmissions and additional retransmission opportunities; intermediate values provide a compromise. These results show that AA-HMARL can adjust its operating behavior to different reliability-latency requirements by varying its preference factor.

\begin{figure*}[!t]
\centering
\renewcommand{\thesubfigure}{\alph{subfigure}}
\subfloat[Problem 1\label{fig:latency:p1}]{
  \includegraphics[width=0.27\textwidth]{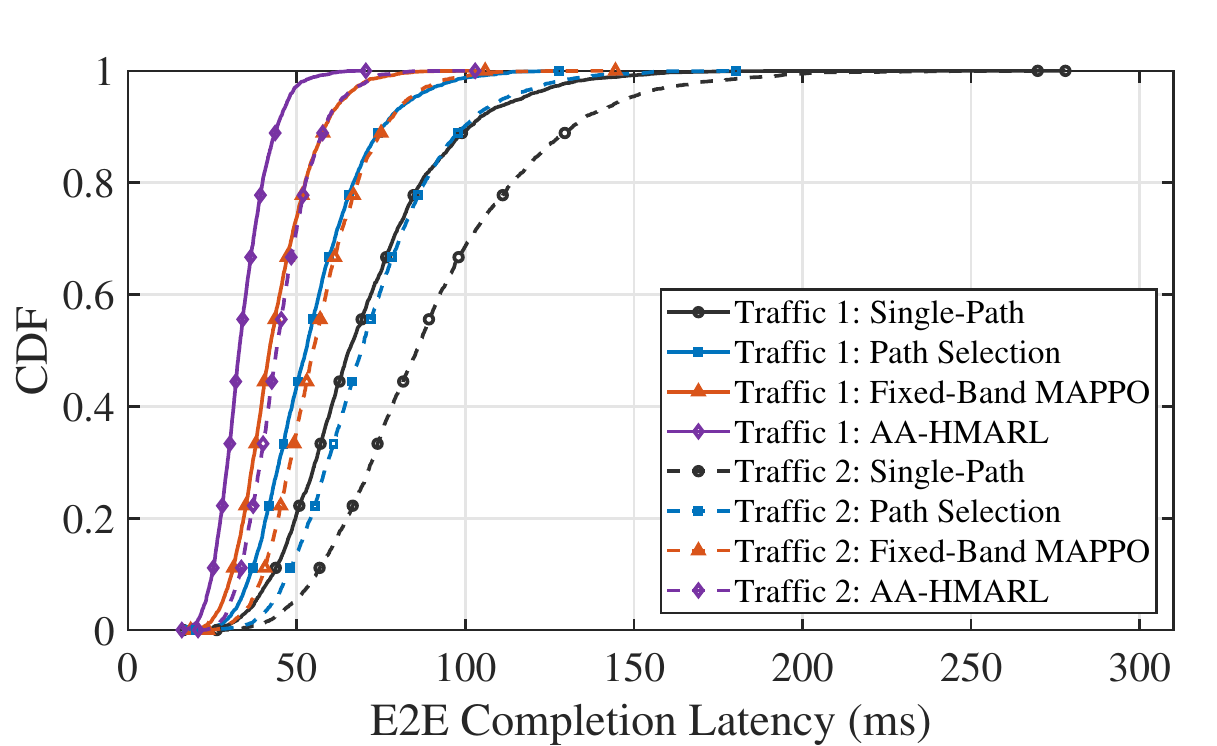}}\hfill
\subfloat[Problem 2\label{fig:latency:p2}]{
  \includegraphics[width=0.27\textwidth]{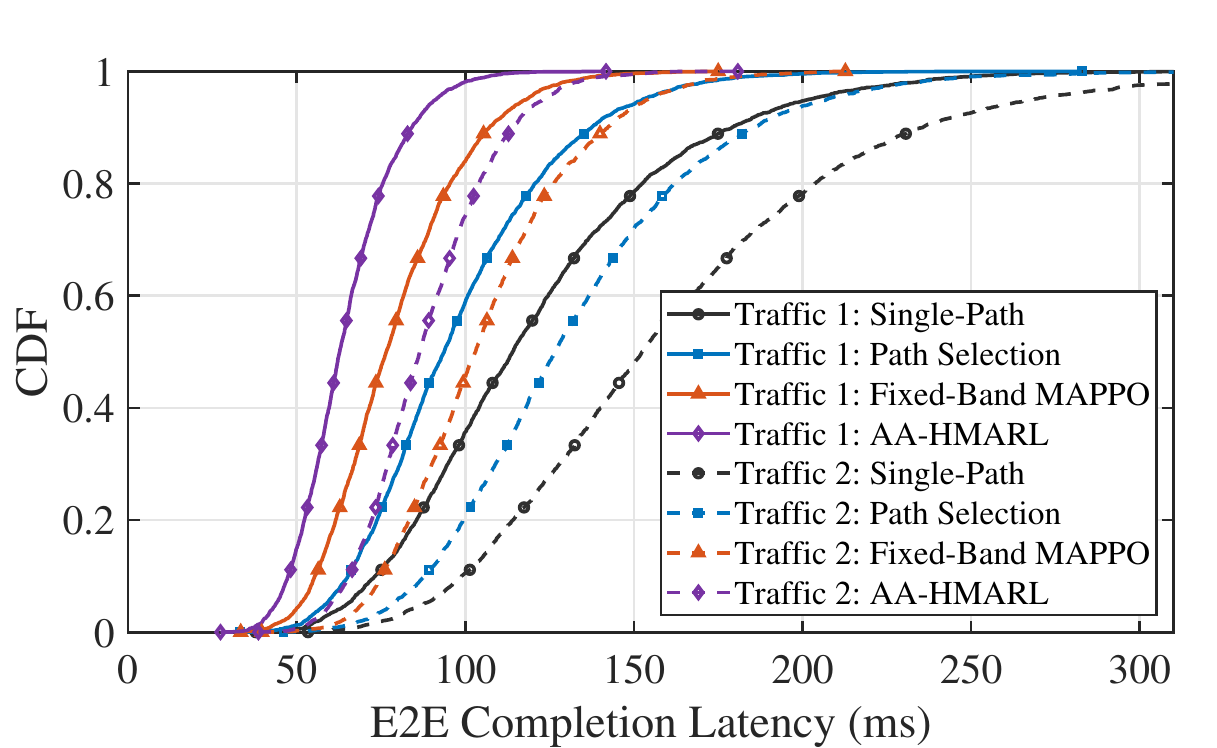}}\hfill
\subfloat[Problem 3\label{fig:latency:p3}]{
  \includegraphics[width=0.27\textwidth]{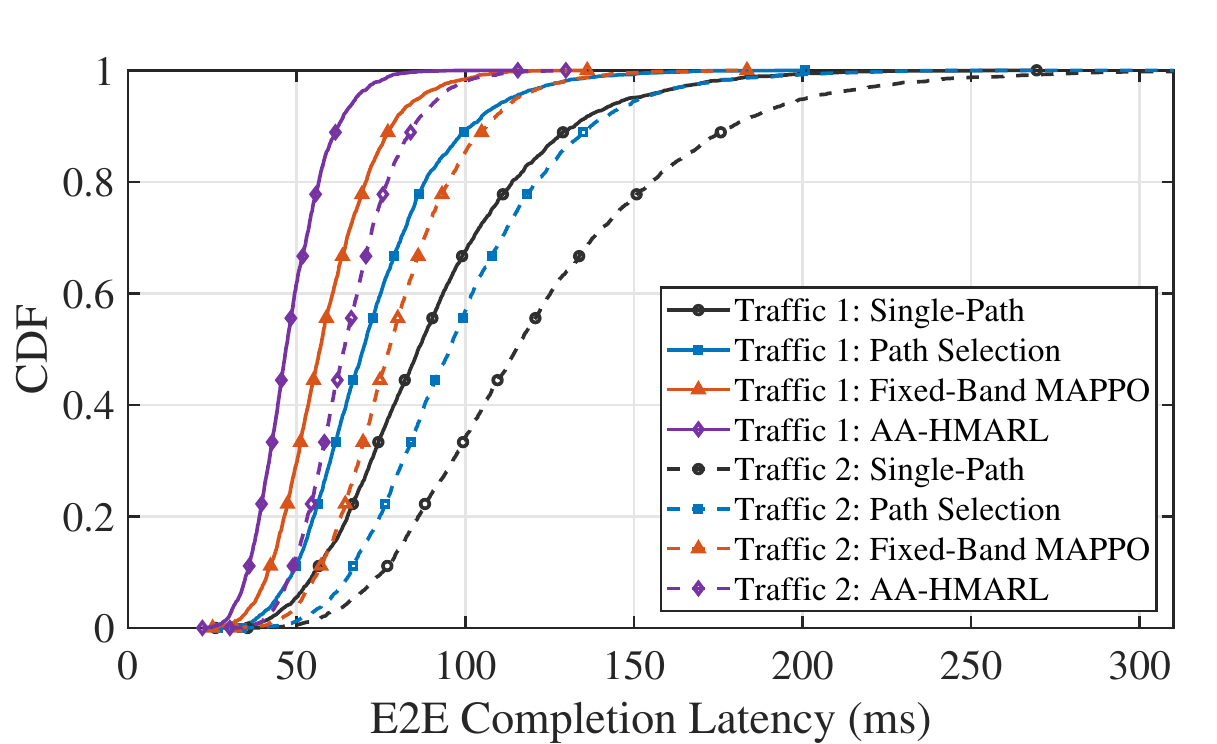}}
  \vspace{-0.2cm}
\captionof{figure}{CDF of E2E completion latency conditioned on successful delivery.}
\vspace{-0.3cm}
\label{fig:latency_cdf}

\subfloat[Problem 1\label{fig:retx:p1}]{
  \includegraphics[width=0.27\textwidth]{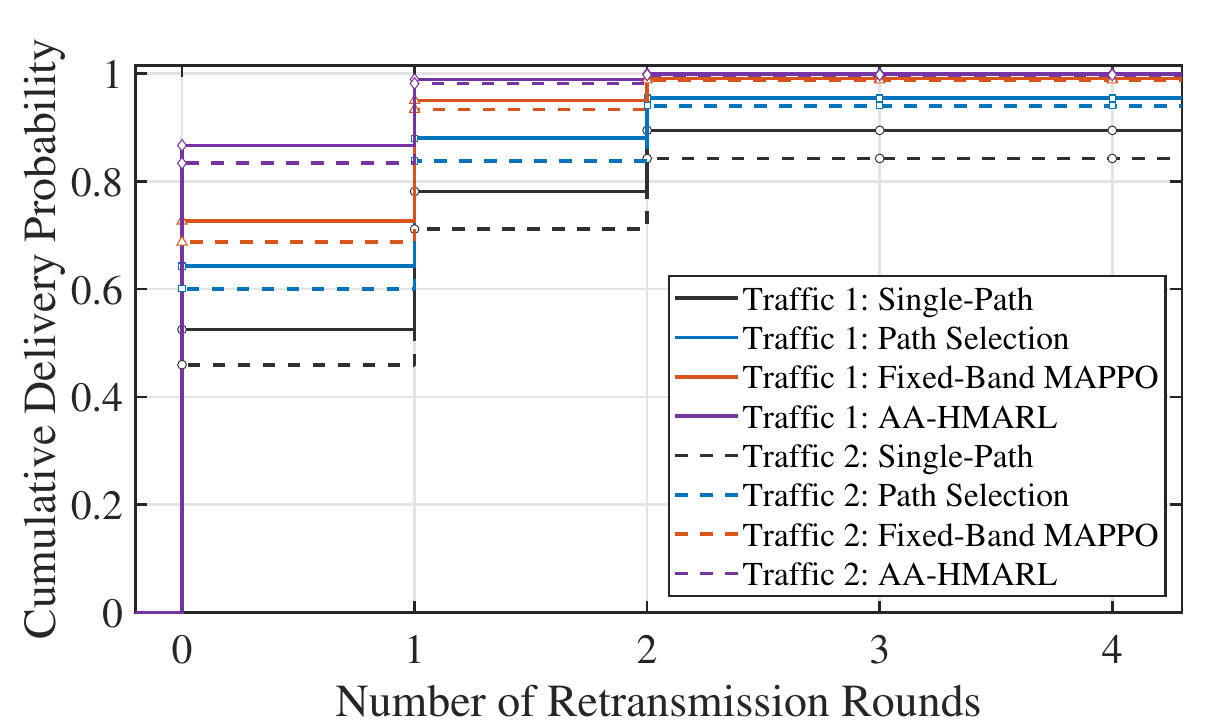}}\hfill
\subfloat[Problem 2\label{fig:retx:p2}]{
  \includegraphics[width=0.27\textwidth]{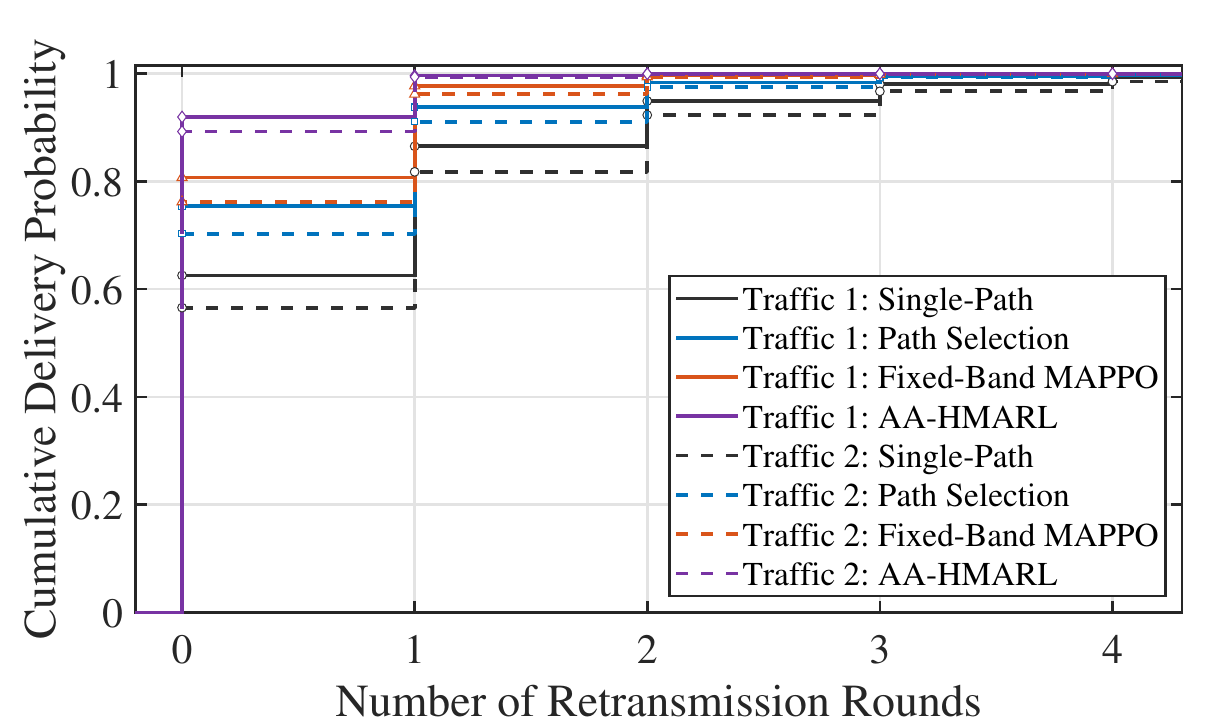}}\hfill
\subfloat[Problem 3\label{fig:retx:p3}]{
  \includegraphics[width=0.27\textwidth]{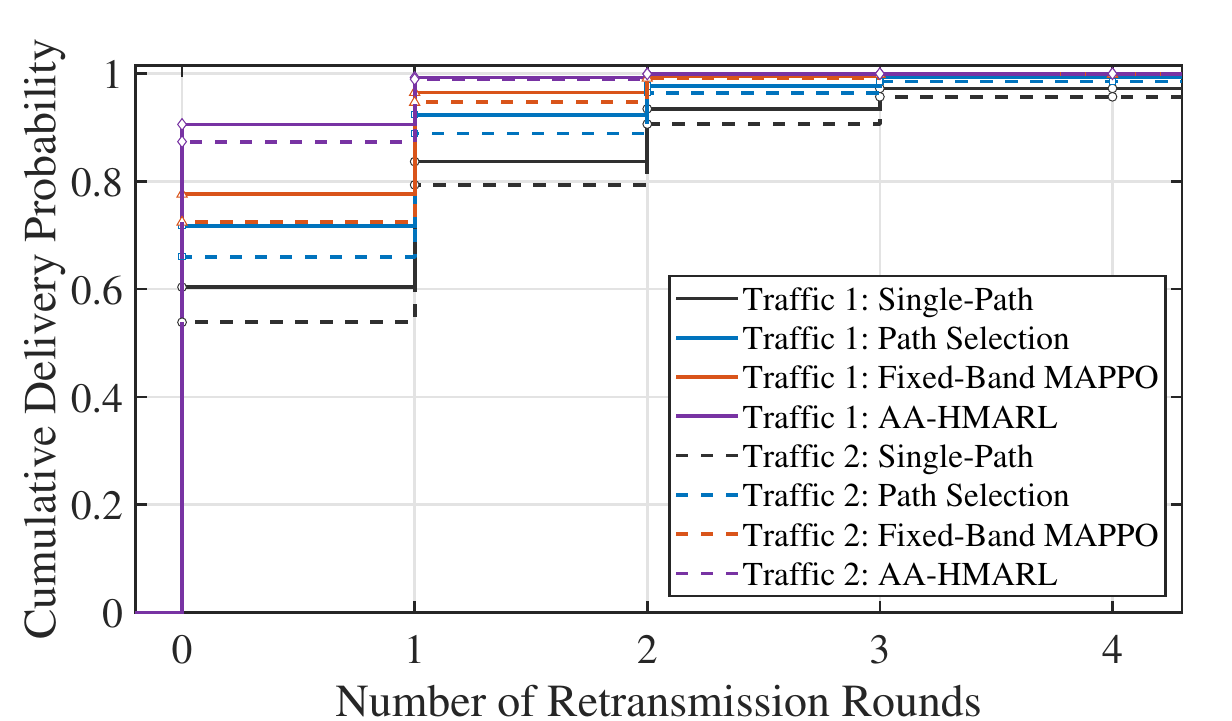}}
    \vspace{-0.2cm}
\captionof{figure}{Cumulative delivery probability in terms of number of sky-link retransmissions.}
\vspace{-0.4cm}
\label{fig:retx_cdf}


\subfloat[ Problem~1\label{fig:aoi:p1}]{
  \includegraphics[width=0.27\textwidth]{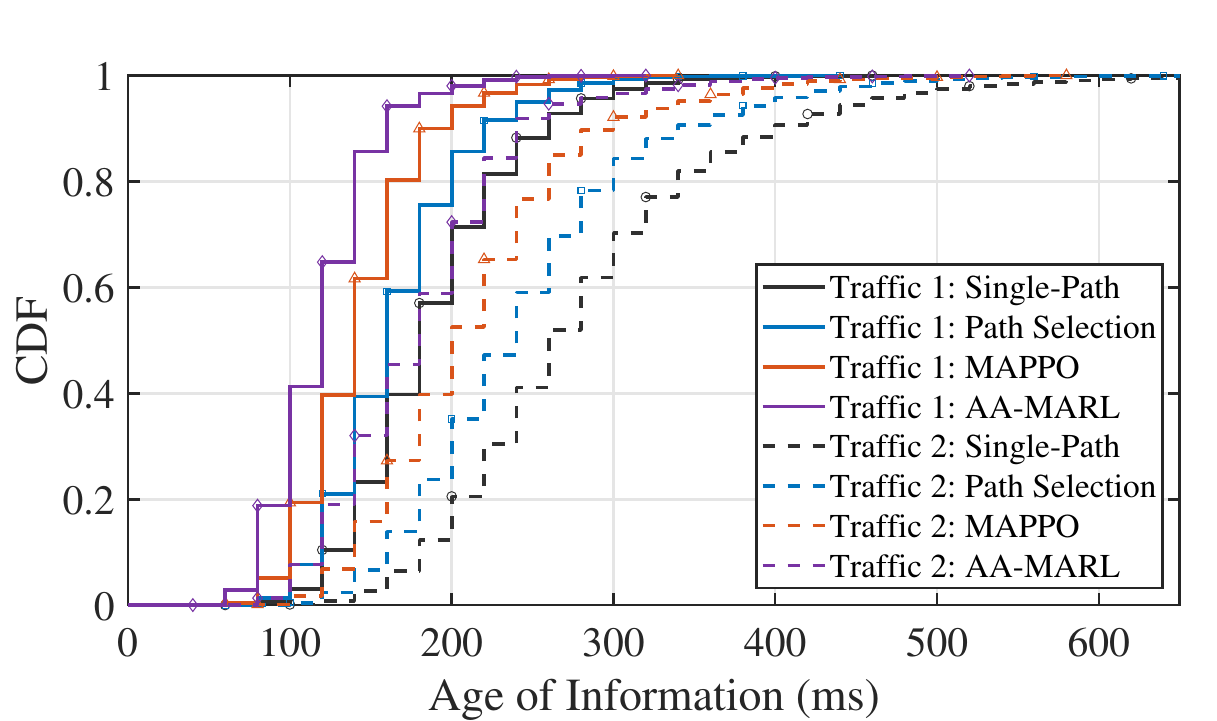}}\hfill
\subfloat[Problem~2\label{fig:aoi:p2}]{
  \includegraphics[width=0.27\textwidth]{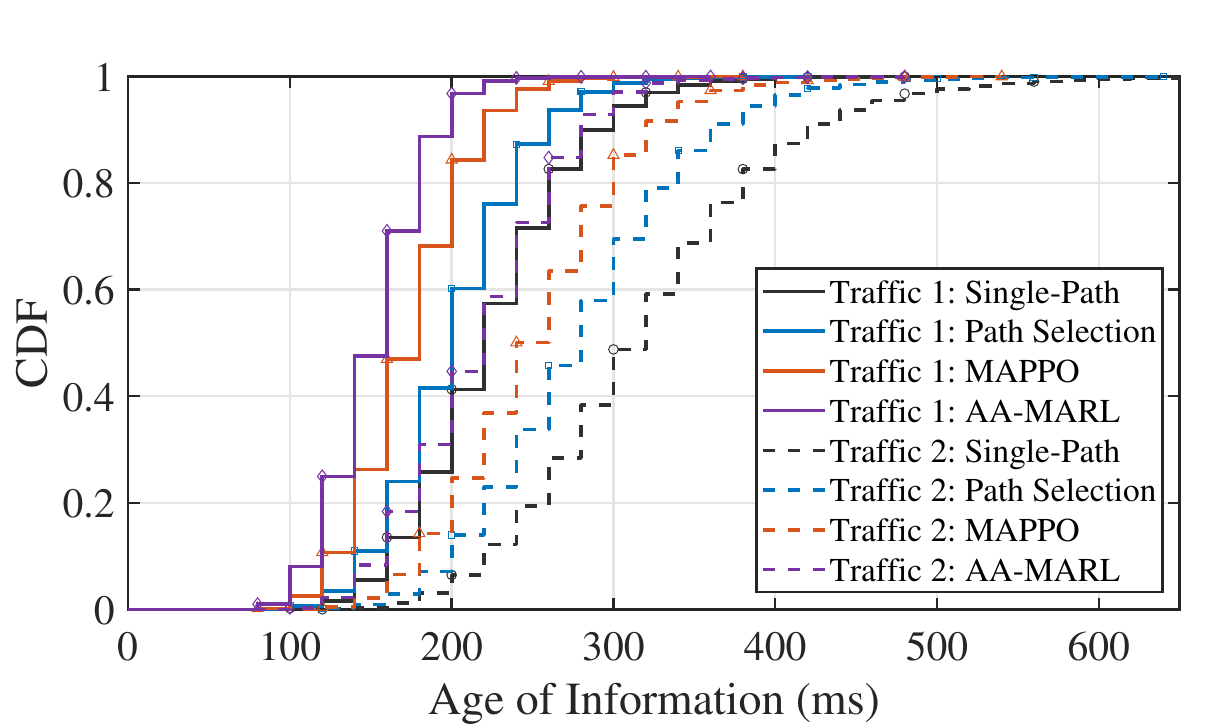}}\hfill
\subfloat[Problem~3\label{fig:aoi:p3}]{
  \includegraphics[width=0.27\textwidth]{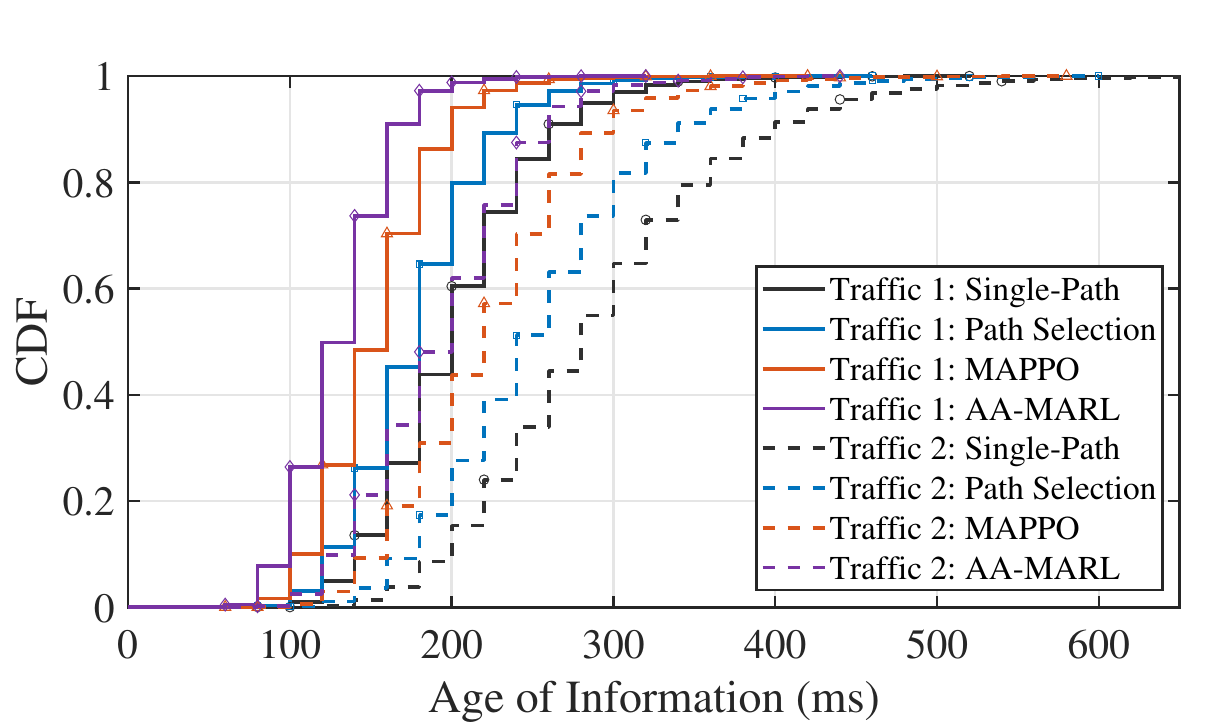}}
    \vspace{-0.2cm}
\captionof{figure}{CDF of receiver-side AoI sampled at fast block boundaries.}
\vspace{-0.6cm}
\label{fig:aoi_cdf}

\end{figure*}


\begin{table}[!t]
\centering
\caption{Average E2E completion latency (ms)}
\vspace{-0.2cm}
\label{tab:average_latency}
\renewcommand{\arraystretch}{1.15}
\setlength{\tabcolsep}{4pt}
\resizebox{\columnwidth}{!}{
\begin{tabular}{lcccccc}
\toprule
\multirow{2}{*}{\textbf{Method}}
& \multicolumn{2}{c}{\textbf{Problem 1}}
& \multicolumn{2}{c}{\textbf{Problem 2}}
& \multicolumn{2}{c}{\textbf{Problem 3}} \\
\cmidrule(lr){2-3}
\cmidrule(lr){4-5}
\cmidrule(lr){6-7}
& Traffic 1 & Traffic 2
& Traffic 1 & Traffic 2
& Traffic 1 & Traffic 2 \\
\midrule
Single-Path
& 69.75 & 90.47
& 120.96 & 162.05
& 90.97 & 122.49 \\

Path Selection
& 54.64 & 72.00
& 98.09 & 132.65
& 72.93 & 99.24 \\

Fixed-Band MAPPO
& 43.61 & 56.73
& 79.36 & 105.93
& 64.53 & 79.78 \\

\textbf{AA-HMARL}
& \textbf{33.88} & \textbf{45.07}
& \textbf{58.82} & \textbf{88.45}
& \textbf{48.06} & \textbf{65.58} \\
\bottomrule
\end{tabular}
}
\end{table}

\begin{table}[!t]
\centering
\caption{Average number of retransmission rounds}
\vspace{-0.2cm}
\label{tab:average_retransmission}
\renewcommand{\arraystretch}{1.15}
\setlength{\tabcolsep}{4pt}
\resizebox{\columnwidth}{!}{
\begin{tabular}{lcccccc}
\toprule
\multirow{2}{*}{\textbf{Method}}
& \multicolumn{2}{c}{\textbf{Problem 1}}
& \multicolumn{2}{c}{\textbf{Problem 2}}
& \multicolumn{2}{c}{\textbf{Problem 3}} \\
\cmidrule(lr){2-3}
\cmidrule(lr){4-5}
\cmidrule(lr){6-7}
& Traffic 1 & Traffic 2
& Traffic 1 & Traffic 2
& Traffic 1 & Traffic 2 \\
\midrule
Single-Path
& 0.54 & 0.61
& 0.55 & 0.68
& 0.56 & 0.66 \\

Path Selection
& 0.40 & 0.47
& 0.32 & 0.41
& 0.36 & 0.45 \\

Fixed-Band MAPPO
& 0.31 & 0.36
& 0.22 & 0.28
& 0.26 & 0.33 \\

\textbf{AA-HMARL}
& \textbf{0.14} & \textbf{0.18}
& \textbf{0.08} & \textbf{0.11}
& \textbf{0.10} & \textbf{0.14} \\
\bottomrule
\end{tabular}
}
\end{table}

\begin{table}[!t]
\centering
\caption{Average AoI (ms)}
\vspace{-0.2cm}
\label{tab:average_aoi}
\renewcommand{\arraystretch}{1.15}
\setlength{\tabcolsep}{4pt}
\resizebox{\columnwidth}{!}{
\begin{tabular}{lcccccc}
\toprule
\multirow{2}{*}{\textbf{Method}}
& \multicolumn{2}{c}{\textbf{Problem 1}}
& \multicolumn{2}{c}{\textbf{Problem 2}}
& \multicolumn{2}{c}{\textbf{Problem 3}} \\
\cmidrule(lr){2-3}
\cmidrule(lr){4-5}
\cmidrule(lr){6-7}
& Traffic 1 & Traffic 2
& Traffic 1 & Traffic 2
& Traffic 1 & Traffic 2 \\
\midrule
Single-Path
& 188.39 & 281.90
& 224.43 & 319.72
& 202.38 & 291.02 \\

Path Selection
& 165.84 & 245.56
& 201.46 & 280.85
& 178.30 & 254.50 \\

Fixed-Band MAPPO
& 143.10 & 213.82
& 174.21 & 252.13
& 153.54 & 222.75 \\

\textbf{AA-HMARL}
& \textbf{119.78} & \textbf{181.47}
& \textbf{152.62} & \textbf{218.39}
& \textbf{131.13} & \textbf{194.40} \\
\bottomrule
\end{tabular}
}
\end{table}

We then compare AA-HMARL with the three baselines under Problems~1-3 using completion-latency, delivery-probability, and freshness distributions.

\emph{1) E2E completion latency:} Fig.~\ref{fig:latency_cdf} reports $\Pr\{D_{k,q}\leq x\mid Y_{k,q}=1\}$ for successfully delivered updates, aggregated across vehicles and test realizations. For both traffic types and all three problems, AA-HMARL shows the strongest leftward shift, followed by FB-MAPPO, PS, and SP over most of the distribution. Its lighter latency tail indicates fewer excessively long successful transmissions under adaptive splitting, power control, and band selection. Table~\ref{tab:average_latency} confirms that AA-HMARL has the lowest mean successful-delivery latency across all problems and both traffic types. Under Problem~3, it reduces the FB-MAPPO values from 64.53 to 48.06~ms for Traffic~1 and from 79.78 to 65.58~ms for Traffic~2, corresponding to reductions of 25.5\% and 17.8\%. Problem~1 achieves shorter completion latency than Problem~2, which generally accepts additional delay for successful delivery; Problem~3 provides an intermediate operating point. These conditional latency results must be read alongside Fig.~\ref{fig:retx_cdf}, since low latency alone could otherwise reflect discarded difficult updates.


\emph{2) Transmission reliability and retransmissions:} Fig.~\ref{fig:retx_cdf} gives the probability of completing the entire update within $r$ additional sky-link retransmission rounds: $F_{\mathrm{del},q}(r)
 =\Pr\{Y_{k,q}=1,\,N_{k,q}^{\mathrm{ret}}\leq r\}$,
where $N_{k,q}^{\mathrm{ret}}$ is the additional-round count for a successfully completed update. For multipath updates, it is determined by the last completed active subflow, rather than the sum of all subflow attempts. All generated updates, including failures, form the normalization set. Thus, $r=0$ measures the first-attempt E2E success, and the terminal plateau gives the packet delivery ratio (PDR) under the retransmission budget. Unlike a distribution conditioned on successful delivery, this curve does not need to reach one; a terminal shortfall accounts for updates remaining undelivered under the retransmission budget. AA-HMARL completes updates with higher probability in fewer rounds than the baselines in each panel. Problem~2 generally attains a higher terminal delivery probability than Problem~1, whereas Problem~3 balances early completion and eventual delivery. These trends reflect the combined effects of adaptive sky-link control and selective retries. Table~\ref{tab:average_retransmission} shows the fewest average retransmission rounds for AA-HMARL across both traffic types. Under Problem~3, the FB-MAPPO averages decrease from 0.26 to 0.10 for Traffic~1 and from 0.33 to 0.14 for Traffic~2, with reductions of 61.5\% and 57.6\%. These counts describe successful-delivery efficiency, while the terminal probabilities in Fig.~\ref{fig:retx_cdf} additionally capture delivery failures among all generated updates.



\begin{figure}[!t]
    \centering
    \renewcommand{\thesubfigure}{\alph{subfigure}}\subfloat[Traffic 1\label{fig:aoi-traffic1}]{       \includegraphics[width=0.23\textwidth]{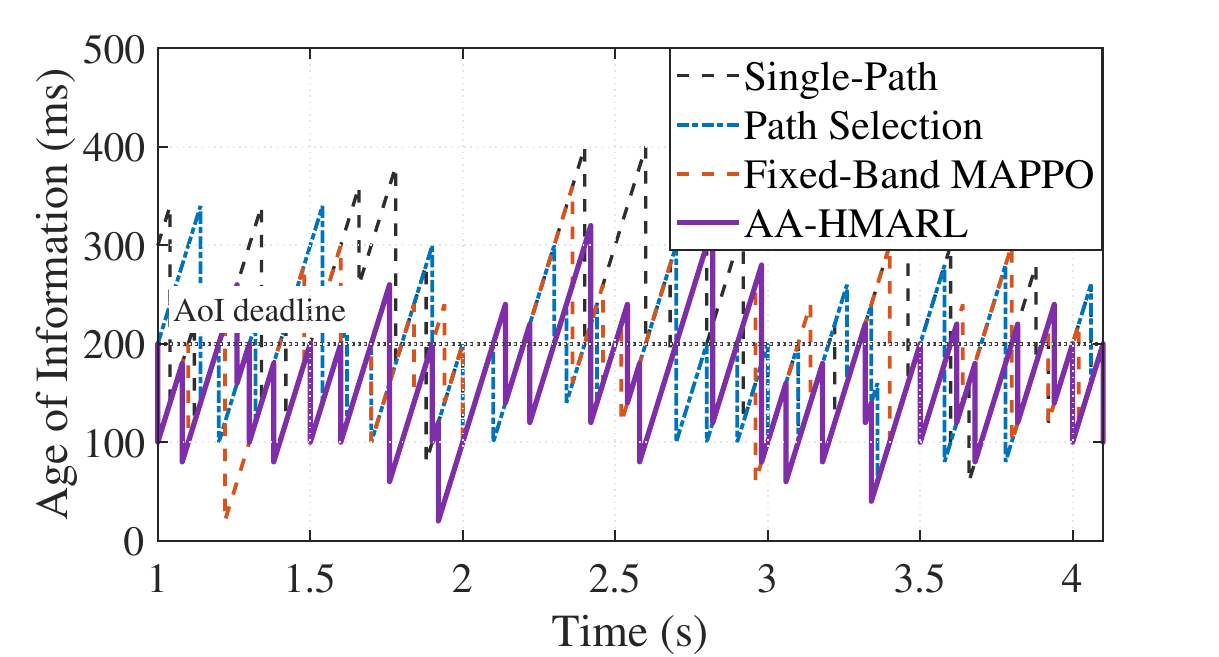}
    }\hfill
    \subfloat[Traffic 2\label{fig:aoi-traffic2}]{        \includegraphics[width=0.23\textwidth]{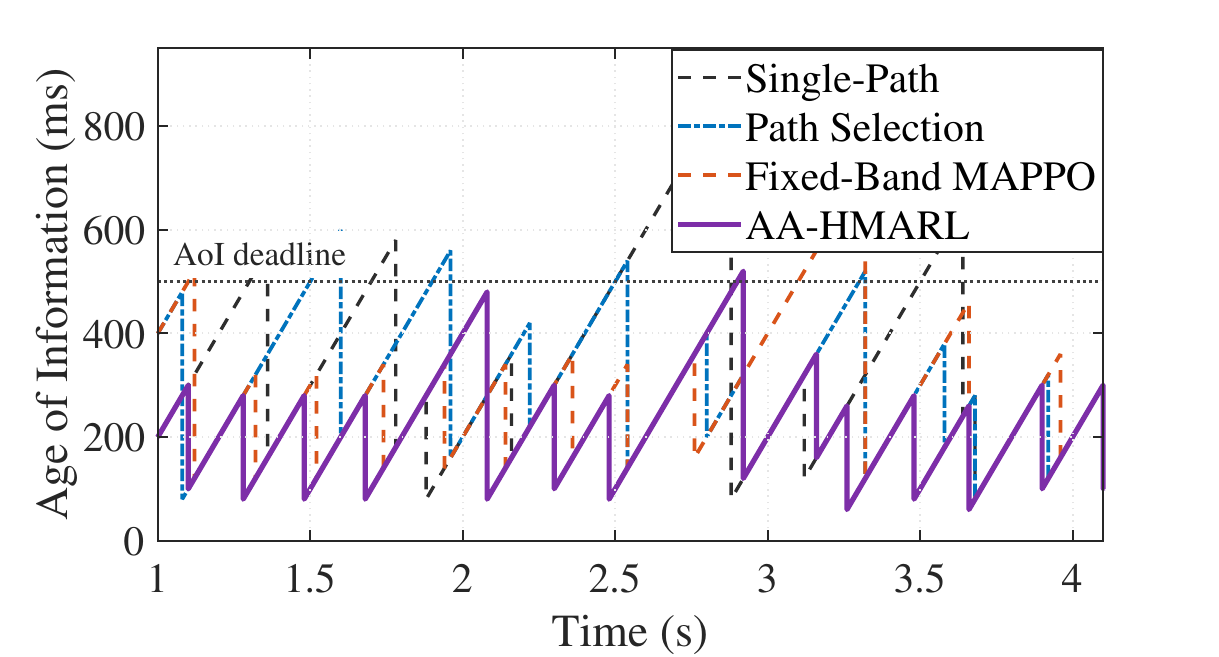}
    }
    \caption{Illustration of AoI evolution for Traffic~1 and Traffic~2.}
    \label{fig:aoi_evolution}
\end{figure}

\emph{3) Information freshness:} Fig.~\ref{fig:aoi_cdf} plots receiver-side AoI sampled at all observed fast-block boundaries, retaining periods of unsuccessful delivery and continued aging. Its step-like curves follow from the discrete AoI recursion. For both traffic types, AA-HMARL places more observations at lower AoI values and suppresses the upper tail relative to the baselines. Problem~1 tends to provide fresher information than Problem~2, with Problem~3 intermediate. Retransmission can improve eventual delivery probability while delaying the next successful AoI refresh. Although AoI is not an independent objective in Problem~3, the freshness-constrained retransmission process and observed AoI state influence scheduling. Table~\ref{tab:average_aoi} confirms that AA-HMARL achieves the lowest average AoI across all three problems and both traffic types. Under Problem~3, it reduces the FB-MAPPO averages from 153.54 to 131.13~ms for Traffic~1 and from 222.75 to 194.40~ms for Traffic~2, corresponding to reductions of 14.6\% and 12.7\%. Fig.~\ref{fig:aoi_evolution} further illustrates the time-domain AoI evolution of the four transmission strategies for Traffic 1 and Traffic 2. In both cases, AoI increases linearly between successful complete-update receptions and drops when a fresher update is fully reassembled at the satellite receiver. As shown in Fig.~\ref{fig:aoi_evolution}(a), Traffic 1 exhibits relatively frequent AoI refreshes owing to its shorter update generation period. In contrast, Fig.~\ref{fig:aoi_evolution}(b) shows larger AoI fluctuations for Traffic 2, reflecting its longer update generation interval and less frequent refresh opportunities. For both traffic types, AA-HMARL generally achieves more frequent effective refreshes and smaller AoI excursions than the baselines, while occasional large peaks under the baselines indicate prolonged intervals without complete update delivery. Nevertheless, the advantage is not uniform at every instant, as a competing strategy may temporarily achieve a lower AoI following a successful reception. Therefore, the long-term AoI distributions and averages provide a more representative performance comparison. These observations also highlight the coupling between E2E subflow reassembly and selective sky-link retransmission, since receiving only part of an update cannot refresh the receiver-side AoI.





\begin{figure}[!t]
    \centering
    \includegraphics[width=0.65\columnwidth]{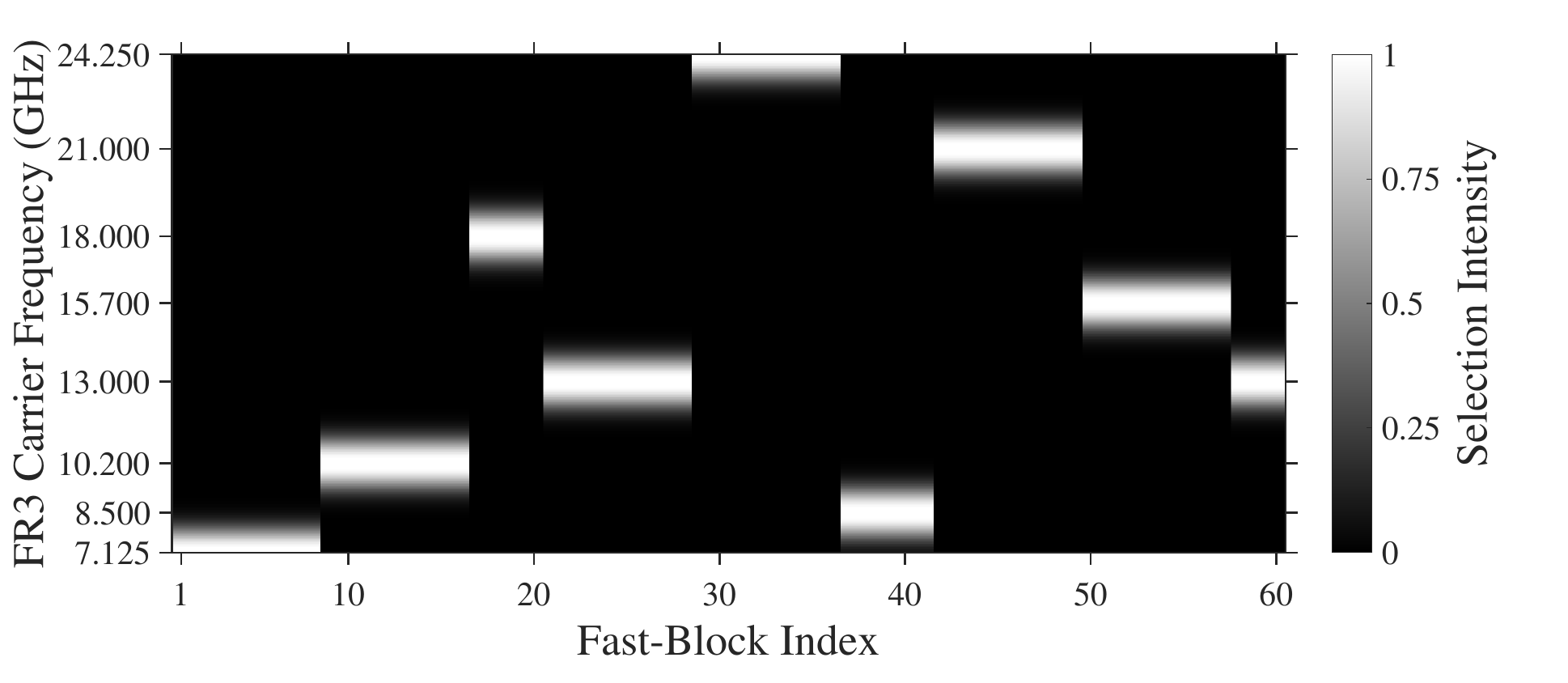}
    \vspace{-0.2cm}
    \caption{Illustration of dynamic FR3 mid-band selection. }
    \label{fig:band_switching}
\end{figure}

Fig.~\ref{fig:band_switching} illustrates an FR3 carrier selection trace over 60 fast transmission blocks (0.6~s). A selection intensity equal to 1 indicates that the optimal FR3 carrier is selected, while an intensity less than 1 indicates the use of a suboptimal carrier when the optimal one is unavailable. The piecewise-constant sequence includes holding intervals shorter than the nominal $N_B$ blocks, which are consistent with Eq. \eqref{eq:option_termination}: QoS degradation or a sufficiently large predicted switching gain may trigger early replanning. Otherwise, the band remains fixed until the next high-level decision. 


\section{Conclusion}
\label{secVII}
This paper studied reliable, low-latency vehicle-UAV-LEO transmission with environment-aware upper-mid-band spectrum agility. An AA-HMARL framework is proposed to coordinate band selection, traffic splitting, and power allocation through an agentic supervisor and fast-timescale actors, with freshness-limited selective retransmission. Under the joint formulation in the simulated scenarios, it reduces mean successful-delivery latency by 17.8-25.5\%, mean retry counts among successfully delivered updates by 57.6-61.5\%, and mean AoI by 12.7-14.6\% relative to fixed-band MAPPO across both traffic types. These results support adjustable reliability-latency operation. Future work will address larger networks, imperfect observations, and practical signaling and reconfiguration costs.

\bibliographystyle{IEEEtran}
\bibliography{ref}

@techreport{3gpp.38.885,
  author       = {{3GPP}},
  title        = {Study on {NR} Vehicle-to-Everything ({V2X})},
  institution  = {3rd Generation Partnership Project (3GPP)},
  type         = {3GPP TR},
  number       = {38.885},
  year         = {2019},
  month        = {Mar}
}

@article{11297179,
  author       = {Bazzi, Ahmad and Bomfin, Roberto and Mezzavilla, Marco and Rangan, Sundeep and Rappaport, Theodore S. and Chafii, Marwa},
  title        = {Upper Mid-Band Spectrum for {6G}: Vision, Opportunity and Challenges},
  journal      = {{IEEE} Commun. Mag.},
  number       = {1},
  volume       = {64},
  pages        = {206--212},
  year         = {2026},
  doi          = {10.1109/MCOM.001.2500101}
}

@article{rappaport2025spectrum,
  author       = {Rappaport, Theodore S. and Humphreys, Todd E. and Nie, Shuai},
  title        = {Spectrum opportunities for the wireless future: from direct-to-device satellite applications to {6G} cellular},
  journal      = {npj Wireless Technol.},
  volume       = {1},
  year         = {2025},
  note         = {Art. no. 8},
  doi          = {10.1038/s44459-025-00008-9}
}

@article{10879563,
  author       = {Zhang, Di and Peng, Yuanyuan and Ke, Ying and Song, Jinpeng and Wang, Shuai and Pan, Gaofeng and An, Jianping},
  title        = {Age of Information Variation of {LEO} Satellite-Terrestrial Uplink Transmissions},
  journal      = {{IEEE} Trans. Veh. Technol.},
  number       = {6},
  volume       = {74},
  pages        = {9645--9655},
  year         = {2025},
  doi          = {10.1109/TVT.2025.3540572}
}

@article{11169451,
  author       = {Tong, Xiaolu and Shi, Yan and Xu, Yaqi and Chen, Shanzhi and Ge, Yuming},
  title        = {{AoI} and {TTC} Based Resource Allocation in {C-V2X} Sidelink via Multi-Agent Reinforcement Learning},
  journal      = {China Commun.},
  number       = {8},
  volume       = {22},
  pages        = {281--297},
  year         = {2025},
  doi          = {10.23919/JCC.fa.2024-0506.202508}
}

@article{11180797,
  author       = {Liao, Mingjun and Zhang, Puning and Huang, Haiyun and Wang, Ruyan and Yang, Junyan and Yang, Zhigang},
  title        = {Multihop Age-of-Information-Enhanced Routing Mechanism for Delay-Sensitive Service in the {LEO} Satellite Networks},
  journal      = {{IEEE} Internet Things J.},
  number       = {23},
  volume       = {12},
  pages        = {51168--51181},
  year         = {2025},
  doi          = {10.1109/JIOT.2025.3612043}
}

@article{10689365,
  author       = {Xu, Yue and Zheng, Linjiang and Wu, Xiao and Tang, Yi and Liu, Weining and Sun, Dihua},
  title        = {Joint Resource Allocation for {UAV}-Assisted {V2X} Communication With Mean Field Multi-Agent Reinforcement Learning},
  journal      = {{IEEE} Trans. Veh. Technol.},
  number       = {1},
  volume       = {74},
  pages        = {1209--1223},
  year         = {2025},
  doi          = {10.1109/TVT.2024.3466116}
}

@article{10977731,
  author       = {Prabhashana, Sasinda C. and Huynh, Dang Van and Singh, Keshav and Zepernick, Hans-Jürgen and Dobre, Octavia A. and Shin, Hyundong and Duong, Trung Q.},
  title        = {Machine Learning-Based Resource Allocation in {6G} Integrated Space and Terrestrial Networks-Aided Intelligent Autonomous Transportation},
  journal      = {{IEEE} Trans. Intell. Transp. Syst.},
  number       = {10},
  volume       = {26},
  pages        = {17750--17762},
  year         = {2025},
  doi          = {10.1109/TITS.2025.3540530}
}

@article{11124240,
  author       = {Tesfaw, Belayneh Abebe and Juang, Rong-Terng and Tarekegn, Getaneh Berie and Kabore, Wendenda Nathanael and Tsai, Ming-Cheng},
  title        = {Joint {UAV} {3-D} Trajectory and Resource Allocation for Integrated {LEO} Satellite and Multi-{UAV}-Enabled Marine {IoT} Networks: A Federated Multiagent Deep Reinforcement Learning Approach},
  journal      = {{IEEE} Internet Things J.},
  number       = {21},
  volume       = {12},
  pages        = {45076--45093},
  year         = {2025},
  doi          = {10.1109/JIOT.2025.3598332}
}

@misc{11550346,
  author       = {Zhang, Linghao and Zhao, Haitao and Xu, Bo and Zhu, Hongbo and Wang, Xianbin},
  title        = {Agentic {AI} for {SAGIN} Resource Management: Semantic Awareness, Orchestration, and Optimization},
  year         = {2026},
  note         = {arXiv:2603.16458},
  doi          = {10.48550/arXiv.2603.16458}
}

@article{juan2018aatr,
  author       = {Juan, Jos{\'e} Miguel and Sanz, Jaume and Rovira-Garcia, Adri{\`a} and Gonz{\'a}lez-Casado, Guillermo and Ib{\'a}{\~n}ez, D. and {Orus Perez}, R.},
  title        = {{AATR} an ionospheric activity indicator specifically based on {GNSS} measurements},
  journal      = {J. Space Weather Space Clim.},
  volume       = {8},
  year         = {2018},
  note         = {Art. no. A14},
  doi          = {10.1051/swsc/2017044}
}

@article{pi1997monitoring,
  author       = {Pi, X. and Mannucci, A. J. and Lindqwister, U. J. and Ho, C. M.},
  title        = {Monitoring of global ionospheric irregularities using the worldwide {GPS} network},
  journal      = {Geophys. Res. Lett.},
  number       = {18},
  volume       = {24},
  pages        = {2283--2286},
  year         = {1997},
  doi          = {10.1029/97GL02273}
}

@article{mahboob2024ai_ntn,
  author       = {Mahboob, Shadab and Liu, Lingjia},
  title        = {Revolutionizing Future Connectivity: A Contemporary Survey on {AI}-Empowered Satellite-Based Non-Terrestrial Networks in {6G}},
  journal      = {{IEEE} Commun. Surveys Tuts.},
  number       = {2},
  volume       = {26},
  pages        = {1279--1321},
  year         = {2024},
  doi          = {10.1109/COMST.2023.3347145}
}

@inproceedings{cui2025fr3,
  author       = {Cui, Zhuangzhuang and Zhang, Peize and Pollin, Sofie},
  title        = {{6G} Wireless Communications in 7--24 {GHz} Band: Opportunities, Techniques, and Challenges},
  booktitle    = {Proc. {IEEE} Int. Symp. Dyn. Spectr. Access Netw. (DySPAN)},
  pages        = {1--8},
  year         = {2025},
  doi          = {10.1109/DySPAN64764.2025.11115934}
}

@article{914515,
  author       = {Krunz, M. M. and Kim, Jeong Geun},
  title        = {Fluid Analysis of Delay and Packet Discard Performance for {QoS} Support in Wireless Networks},
  journal      = {{IEEE} J. Sel. Areas Commun.},
  number       = {2},
  volume       = {19},
  pages        = {384--395},
  year         = {2001},
  doi          = {10.1109/49.914515}
}

@misc{wang2025environment,
  author       = {Wang, Houtianfu and Dong, Haofan and Cai, Hanlin and Akan, Ozgur B},
  title        = {Environment-to-Link {ISAC} with Space-Weather Sensing for {Ka}-Band {LEO} Downlinks},
  year         = {2026},
  note         = {arXiv:2601.00820},
  doi          = {10.48550/arXiv.2601.00820}
}

@misc{gong2026starmind,
  author       = {Gong, Zisheng and Liu, Zhaoyu and Wang, Houtianfu and Baydas, O Tansel and Cao, Liu and Xu, Xia and Akan, Ozgur B},
  title        = {{StarMind}: Challenges and Opportunities in Orbital Computing},
  year         = {2026},
  note         = {arXiv:2610.26916},
  doi          = {10.48550/arXiv.2610.26916}
}

@article{10645749,
  author       = {Nguyen-Kha, Hung and Nguyen Ha, Vu and Lagunas, Eva and Chatzinotas, Symeon and Grotz, Joel},
  title        = {Joint Two-Tier User Association and Resource Management for Integrated Satellite-Terrestrial Networks},
  journal      = {{IEEE} Trans. Wireless Commun.},
  number       = {11},
  volume       = {23},
  pages        = {16648--16665},
  year         = {2024},
  doi          = {10.1109/TWC.2024.3445266}
}

@article{10937131,
  author       = {Cheng, Lei and Li, Xiaoqian and Feng, Gang and Peng, Youkun and Qin, Shuang and Quek, Tony Q.S.},
  title        = {Cooperative Transmission for Space-Air-Ground Integrated Networks: A Multi-Agent Cooperation Method},
  journal      = {{IEEE} Trans. Veh. Technol.},
  number       = {8},
  volume       = {74},
  pages        = {12879--12894},
  year         = {2025},
  doi          = {10.1109/TVT.2025.3553625}
}

@article{10990315,
  author       = {Minani, Frodouard and Kobayashi, Makoto and Fujihashi, Takuya and Alim, Md. Abdul and Saruwatari, Shunsuke and Nishi, Masahiro and Watanabe, Takashi},
  title        = {Channel Prediction and Fair Resource Allocation for {NTN} Uplinks by {LSTM} and Deep Reinforcement Learning},
  journal      = {{IEEE} Trans. Wireless Commun.},
  number       = {10},
  volume       = {24},
  pages        = {8311--8330},
  year         = {2025},
  doi          = {10.1109/TWC.2025.3565774}
}

@article{SUTTON1999181,
  author       = {Richard S. Sutton and Doina Precup and Satinder Singh},
  title        = {Between {MDPs} and semi-{MDPs}: A framework for temporal abstraction in reinforcement learning},
  journal      = {Artif. Intell.},
  number       = {1--2},
  volume       = {112},
  pages        = {181--211},
  year         = {1999},
  doi          = {10.1016/S0004-3702(99)00052-1},
  issn         = {0004-3702}
}

@inproceedings{pmlr-v48-martins16,
  author       = {Martins, Andre and Astudillo, Ramon},
  title        = {From Softmax to Sparsemax: A Sparse Model of Attention and Multi-Label Classification},
  booktitle    = {Proc. 33rd Int. Conf. Mach. Learn. ({ICML})},
  series       = {Proc. Mach. Learn. Res.},
  volume       = {48},
  pages        = {1614--1623},
  year         = {2016},
  month        = jun,
  publisher    = {PMLR}
}

@article{10438999,
  author       = {Liu, Yalin and Dai, Hong-Ning and Wang, Qubeijian and Pandey, Om Jee and Fu, Yaru and Zhang, Ning and Niyato, Dusit and Lee, Chi Chung},
  title        = {Space-Air-Ground Integrated Networks: Spherical Stochastic Geometry-Based Uplink Connectivity Analysis},
  journal      = {{IEEE} J. Sel. Areas Commun.},
  number       = {5},
  volume       = {42},
  pages        = {1387--1402},
  year         = {2024}
}

@article{10795214,
  author       = {Tun, Yan Kyaw and Dán, György and Park, Yu Min and Hong, Choong Seon},
  title        = {Joint {UAV} Deployment and Resource Allocation in {THz}-Assisted {MEC}-Enabled Integrated Space-Air-Ground Networks},
  journal      = {{IEEE} Trans. Mobile Comput.},
  number       = {5},
  volume       = {24},
  pages        = {3794--3808},
  year         = {2025},
  doi          = {10.1109/TMC.2024.3516655}
}

@techreport{3gpp38300,
  author       = {{3GPP}},
  title        = {{NR}; {NR} and {NG-RAN} Overall description; Stage-2},
  institution  = {3rd Generation Partnership Project},
  type         = {3GPP TS},
  number       = {38.300},
  year         = {2024}
}

@techreport{3gpp_rel17_ntn,
  author       = {{3GPP}},
  title        = {Release 17 Description; Summary of Rel-17 Work Items},
  institution  = {{3rd Generation Partnership Project (3GPP)}},
  type         = {3GPP TR},
  number       = {21.917},
  year         = {2023},
  month        = jan
}

@techreport{3gpp_rel18_ntn,
  author       = {{3GPP}},
  title        = {Release 18 Description; Summary of Rel-18 Work Items},
  institution  = {{3rd Generation Partnership Project (3GPP)}},
  type         = {3GPP TR},
  number       = {21.918},
  year         = {2025},
  month        = apr
}

@misc{deusser2026evaluating,
  title  = {Evaluating and Benchmarking the
            {System One} Model {Jev}},
  author = {Deu{\ss}er, Tobias and
            Ramamurthy, Rajkumar and
            Sparrenberg, Lorenz and
            Sifa, Rafet},
  year   = {2026},
  note   = {arXiv:2609.37647v2},
  doi    = {10.48550/arXiv.2609.37647}
}

@inproceedings{NEURIPS2024_fa76985f,
  author    = {Zhang, Lijun and others},
  title     = {Scalable Constrained Policy
               Optimization for Safe Multi-Agent
               Reinforcement Learning},
  booktitle = {Adv. Neural Inf. Process. Syst.
               ({NeurIPS})},
  volume    = {37},
  pages     = {138698--138730},
  year      = {2024},
  doi       = {10.52202/079017-4400}
}

@inproceedings{10622306,
  author    = {Tian, Yu and Alhammadi, Ahmed and
               He, Jiguang and Fakhreddine, Aymen and
               Bader, Faouzi},
  title     = {Fairness-Driven Optimization of
               {RIS}-Augmented {5G} Networks for
               Seamless {3D} {UAV} Connectivity
               Using {DRL} Algorithms},
  booktitle = {Proc. {IEEE} Int. Conf. Commun. ({ICC})},
  pages     = {5009--5014},
  year      = {2024},
  doi       = {10.1109/ICC51166.2024.10622306}
}
\end{document}